\documentclass[modern]{aastex701}
\usepackage{CJK}
\def\fe{[\ion{Fe}{2}]}
\def\si{[\ion{Si}{1}]}
\def\cen{$(23^{\rm h}23^{\rm m}27\fs77,\ +58\arcdeg48\arcmin49\farcs4)$}
\def\unit{erg cm$^{-2}$ s$^{-1}$}
\def\kms{~km~s$^{-1}$}
\def\hst{\textrm{HST}}
\def\jwst{\textrm{JWST}}
\def\img{deep \fe+\si\ image}
\def\simlt{\lower.5ex\hbox{$\; \buildrel < \over \sim \;$}}
\def\simgt{\lower.5ex\hbox{$\; \buildrel > \over \sim \;$}}

\begin{document}
\begin{CJK*}{UTF8}{mj}
\title{A Comparative Study of the Supernova Remnant Cassiopeia A from 2013--2020 Deep \fe+\si\ Images}

\correspondingauthor{Bon-Chul Koo}

\author[0009-0005-6640-4715]{Seung-Hoon Jung (정승훈)}
\affiliation{Department of Physics and Astronomy, Seoul National University, Seoul 08826, Republic of Korea}
\email{femtomino@snu.ac.kr}
\author[0000-0002-2755-1879]{Bon-Chul Koo (구본철)}
\affiliation{Department of Physics and Astronomy, Seoul National University, Seoul 08826, Republic of Korea}
\email[show]{bckoo@snu.ac.kr}
\author[0000-0003-3277-2147]{Yong-Hyun Lee (이용현)}
\affiliation{Korea Astronomy and Space Science Institute, Daejeon 305-348, Republic of Korea}
\affiliation{Samsung SDS, Olympic-ro 35-gil 125, Seoul, Republic of Korea}
\email{yhlee.astro@gmail.com}
\author[0000-0001-9263-3275]{Hyun-Jeong Kim (김현정)}
\affiliation{Korea Astronomy and Space Science Institute, Daejeon 305-348, Republic of Korea}
\email{hjkim@kasi.re.kr}
\author[0000-0003-0894-7824]{Jae-Joon Lee (이재준)}
\affiliation{Korea Astronomy and Space Science Institute, Daejeon 305-348, Republic of Korea}
\email{lee.j.joon@gmail.com}

\begin{abstract}
We present a comparative analysis of supernova remnant Cassiopeia A based on two deep, narrow-band images covering the \fe\ 1.644~\micron\ + \si\ 1.645~\micron\ lines obtained in 2013 and 2020 with the same instruments on the UKIRT 3.8~m telescope. The identical setup and observing procedure allow for direct, accurate measurements of morphological and kinematic changes over a seven-year baseline. We identified 263 compact knots in the 2020 image and, through comparison with the 2013 catalog of \citet{koo18}, classified them into quasi-stationary circumstellar knots and fast-moving knots (FMKs) of supernova ejecta. The FMKs show significant flux fluctuations, and many of those detected in 2013 are absent in the 2020 image. Proper-motion measurements derived from cross-correlation analysis indicate that most FMKs follow nearly ballistic expansion, whereas some, particularly those just beyond the eastern Fe-rich, X-ray emitting ejecta region, exhibit noticeable deceleration. The proper motions of the main ejecta shell were also measured and modeled as a uniformly expanding shell with a systemic motion, which reproduces the observed geometric and kinematic asymmetries of the remnant.
\end{abstract}

\section{Introduction \label{sec:intro}}
Cassiopeia~A (Cas~A) is a Galactic Type IIb supernova remnant (SNR) \citep{krause08}. Being both young \citep[$\sim350$~yr;][]{thorstensen01,reed95} and nearby  \citep[3.4~kpc;][]{reed95,alarie14,neumann24}, it offers a unique opportunity to study the details of supernova (SN) explosion dynamics and the interaction of SN ejecta with the circumstellar medium (CSM) expelled by the progenitor star prior to explosion. 

The SN ejecta in Cas~A are distributed in three distinct morphological and physical regimes. Closest to the explosion site lies a volume of unshocked, freely expanding ejecta that retains the imprint of the asymmetric and multi-dimensional explosion. Because of their faintness, the spatial and kinematic properties of these unshocked ejecta have been only marginally characterized \citep{isensee10,grefenstette14,grefenstette17,delaney14,milisavljevic15,koo18}. Recent James Webb Space Telescope (\jwst) observations, however, have uncovered a striking weblike network of unshocked oxygen filaments, highlighting the turbulent mixing and hydrodynamical instabilities that developed shortly after core collapse \citep{milisavljevic24,orlando25a}. 

Surrounding the unshocked ejecta is the bright `main ejecta shell', a non-uniform and clumpy shell of radius $\sim1\farcm7$ (1.7~pc at 3.4~kpc), where intermediate-mass elements have been heated by the reverse shock --- the shock driven by the high-pressure of the shocked CSM \citep{mckee74} --- and now dominates the SNR's optical, infrared, and X-ray emission. Three-dimensional Doppler reconstruction shows that the ejecta in the main shell are organized into several large ring-like structures projected across the spherical shell, features that may be associated with post-explosion input of energy from plumes of radioactive $^{56}$Ni-rich ejecta \citep{lawrence95,delaney10,milisavljevic13,orlando21}. 

Beyond the main shell lie numerous small, dense ejecta `knots'  that have overrun the SN blast wave. These outlying fast-moving knots (FMKs) propagate at significantly higher velocities than the bulk of the O- and S-rich ejecta in the main shell, reaching $\gtrsim15,000$\kms. Their composition and kinematics indicate that they trace high-velocity outflows of material from deep layers of the progenitor star \citep{fesen01a,fesen06,hammell08,koo23}. Such knots are thought to arise from hydrodynamic instabilities developed during the SN explosion,  with radiative cooling playing a crucial role \citep{orlando16,orlando21,orlando25b}, or possibly from jet-like outflows launched by the newly formed neutron star \citep[e.g.,][]{burrows05,burrows21,janka22}. Together, these three components --- unshocked interior ejecta, shocked ejecta confined to the bright shell, and outlying FMKs --- encapsulate the complex hydrodynamics of Cas~A's Type IIb explosion. 

The main ejecta shell and the outlying FMKs, both prominent in the optical, have been central to kinematic studies probing Cas~A's explosion dynamics since the earliest observations \citep{kamper76,vandenbergh83,lawrence95,reed95,thorstensen01,fesen06,milisavljevic13}. The term ``fast-moving knots'' was first introduced in early optical studies to describe the bright filaments and knots in the main ejecta shell \citep[e.g.,][]{vandenbergh70,vandenbergh71}, whereas the large population of the outlying FMKs was recognized only later \citep{fesen01a,fesen06,hammell08}. In this paper, we restrict the term FMKs to those lying beyond the main shell, while referring to those within the shell as ejecta knots in the main shell or simply as main shell ejecta. The FMKs are nearly undecelerated, and their proper motions have been used to determine Cas~A's center of expansion \cen\ and an explosion epoch of 1671.3$\pm$0.9 \citep{thorstensen01}. A more recent Hubble Space Telescope (\hst) study confirmed the explosion date, albeit with larger uncertainties \citep[i.e., 1672$\pm$18;][]{fesen06}. These proper-motion studies also found measurable deceleration of the FMKs, suggesting variations in their local environments and evolutionary histories. 
 
The main ejecta shell exhibits spatially dependent expansion: In the plane of the sky, the western rim expands significantly faster than the eastern side \citep{reed95,thorstensen01,fesen25}. Although the shell appears nearly circular with a radius of $\sim$1\farcm7, the distance from the explosion center,  which is tightly constrained by FMK proper-motion measurements \citep{thorstensen01}, to the western rim is noticeably greater than that to the eastern rim. Along the line-of-sight (LOS), the velocity distribution of the main shell ejecta is further offset by about $+$800\kms\ \citep{reed95,delaney10,milisavljevic13,alarie14}. These spatial and kinematic asymmetries suggest that the SNR is globally asymmetric, likely due to an intrinsically asymmetric explosion, a non-uniform circumstellar environment, or a combination of both.

In this paper, we present a comparative study of Cas~A's kinematics in the main ejecta shell and outer FMKs using two near-infrared (NIR) \fe\ 1.644~\micron\ line images obtained in 2013 and 2020. This forbidden transition of singly ionized iron (Fe$^+$) is prominent in shocked ejecta material \citep{gerardy01,rho03,koo13,koo18}. In addition, being in the NIR, the line suffers much less from interstellar extinction compared to optical lines, making it especially effective for studying heavily obscured SNRs such as Cas~A. 

A detailed \fe\ line study of Cas~A was conducted by \citet{koo18}, who obtained a long-exposure 1.64~\micron\ narrow-band image of Cas~A with the United Kingdom Infrared Telescope (UKIRT) in September 2013. The bandpass also includes the \si\ 1.645~\micron\ line in addition to the \fe\ 1.644~\micron; however, this line is faint and becomes non-negligible only for unshocked ejecta in the inner regions of the remnant (see Section~\ref{ssec:2020}). The image (hereafter the “2013 \img”) therefore primarily traces \fe\ line emitting material.

The image, obtained with an effective net exposure time of 10~hr, provided an unprecedented panoramic view of Cas~A, showing both shocked SN ejecta and shocked CSM at subarcsecond ($\sim$0\farcs7 or 0.012~pc) resolution. From the image, \citet{koo18} identified a few hundred dense knots and classified them into fast-moving SN ejecta knots and slower, nearly stationary, circumstellar knots, known as quasi-stationary flocculi \citep[QSFs;][]{peimbert71,vandenbergh71,kamper76}. However, the classification was based on comparison with the image obtained in August 2008 using the Palomar telescope, which had lower sensitivity and worse seeing. As a result, several dozen knots remained only candidate identifications. The limitations in image quality also made it difficult to investigate the shell expansion. To address these shortcomings, we obtained another \img\ in 2020 using the same instrument and observing procedure (hereafter the ``2020 \img''), enabling a direct and more accurate comparison with the 2013 data.

The paper is organized as follows. Section~\ref{sec:observe} describes the observations and image processing. In Section~\ref{sec:overview}, we present an overview of the 2020 \img, along with a difference image obtained by subtracting the 2013 data. In Section~\ref{sec:catalog}, we introduce the catalog of knots outside the main shell, classified into QSFs and FMKs based on their proper motions. We also summarize changes in the knot population and flux variability since 2013. Section~\ref{sec:decel} presents measurements of FMK proper motions and their deceleration, while Section~\ref{sec:shell} examines the proper motions of the main shell ejecta, emphasizing asymmetries in expansion. Finally, Section~\ref{sec:conclusion} provides a summary and conclusions.

\section{Observation and Data Reduction \label{sec:observe}}

The narrow-band \fe+\si\ 1.64~\micron\ observations, along with the $H$-band observations, were conducted using the Wide-Field Camera (WFCAM) on the UKIRT 3.8~m telescope during July and August 2020. WFCAM is equipped with four Rockwell Hawaii-II HgCdTe infrared focal plane arrays, each with 2048$\times$2048 pixels. A single array provides a field of view (FoV) of 13\farcm65$\times$13\farcm65 with a pixel scale of 0\farcs4. The arrays are arranged in a square pattern, with a gap of 12\farcm83 between them. The observing procedure follows the method described in \citet{koo18}. Since the FoV of a single array is significantly larger than the entire Cas~A, we captured images by alternating between two arrays, which allowed us to use the exposure from one array for flat-fielding and sky subtraction. This resulted in two sky-subtracted images for each set of exposures. A 2$\times$2 microstepping sequence with an interfacing technique has been used to fully sample the point-spread function (PSF). By combining all the images obtained from the two arrays, we created an \fe+\si\ image with an effective net exposure time of 5.8~hours and an $H$-band image with 0.44~hours. This 2020 \img\ has a pixel size of 0\farcs2 and a median PSF of 0\farcs84. The PSF is slightly worse than that (0\farcs7) of the 2013 \img. Astrometric and photometric calibrations were performed as described in \citet{koo18}. The 1$\sigma$ uncertainty in astrometry is 0\farcs125 (or 0.62~pixel) and the flux calibration uncertainty is 5\% (1$\sigma$). The continuum-subtracted image presented in this article has been produced following the procedure in \citet{lee14a}, and has a sensitivity (1$\sigma$) of approximately 3.2$\times 10^{-18}$ \unit\ pixel$^{-1}$. This sensitivity is slightly higher than that (2.6$\times 10^{-18}$ \unit\ pixel$^{-1}$) of the 2013 \img.

As noted in \citet{koo18}, there are some caveats. First, the flux of \fe\ emission features derived from the continuum-subtracted image may be underestimated due to the presence of \fe\ lines in the $H$-band \citep[1.5--1.8~\micron; see Table~1 of][]{koo16}. The scaling factor for this underestimation depends on the electron density ($n_e$) and temperature ($T_e$) of the source, ranging from 1.16 to 1.32 for gas with $n_e=10^3$--$10^5$ cm$^{-3}$ and $T_e=7000$~K. We use a scaling factor of 1.2 in this paper. Another caveat is that the \fe+\si\ filter has a width of $\pm$2600\kms, which means that high-velocity ejecta appear as negative features in our \fe+\si\ image due to the subtraction of the $H$-band image. However, such instrumental artifacts are primarily confined to the northern central region.

\section{Overview of the \fe+\si\ Image \label{sec:overview}}

\subsection{2020 Deep \fe+\si\ Image \label{ssec:2020}}

\begin{figure}[]
\fig{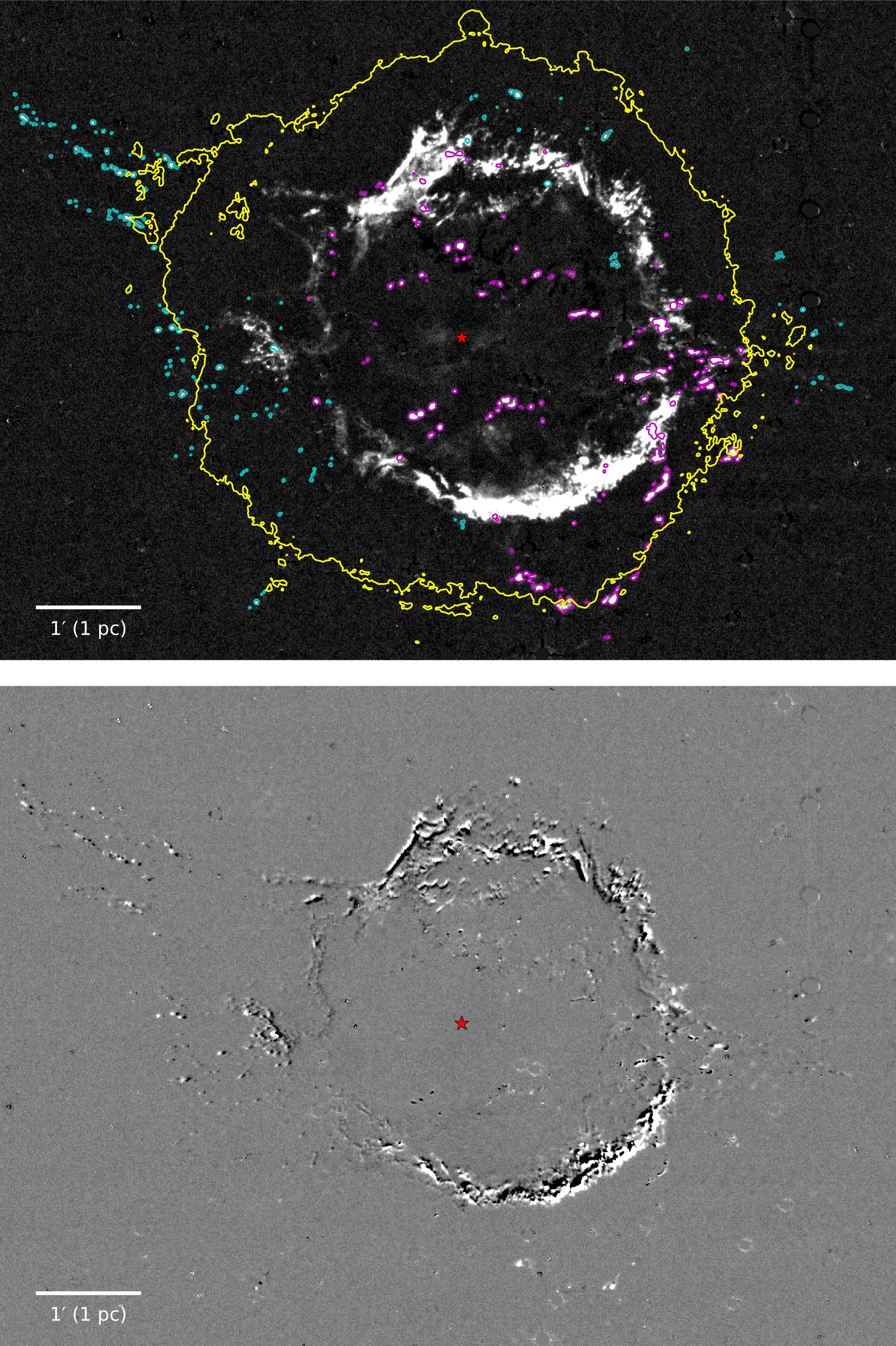}{0.7\textwidth}{}
\caption{Top: The 2020 \img\ of Cas~A obtained from the UKIRT 3.8~m telescope. North is up, and east is to the left. Stellar sources have been removed. The gray scale varies linearly from $-2\times10^{-18}$ to $2\times10^{-17}$ \unit\ pixel$^{-1}$. QSFs and FMKs are marked by magenta and cyan contours, respectively. Yellow contours mark the radio boundary of the SNR from the VLA 6~cm image with a threshold level 0.3~mJy beam$^{-1}$ \citep{delaney04}. The red star represents the explosion center at $(\alpha, \delta)_{\rm J2000} = $ \cen\ \citep{thorstensen01} Bottom: Image difference created by subtracting the 2013 \img\ from the 2020 \img. The gray scale varies linearly from $-2\times10^{-17}$ to $2\times10^{-17}$ \unit\ pixel$^{-1}$. \label{fig:overview}}
\end{figure}

Figure~\ref{fig:overview} presents our 2020 \img\ of the Cas~A. The overall morphology is essentially consistent with that of the 2013 \img\ presented in \citet{koo18} as expected, since no significant morphological evolution is expected over this timescale. This narrow-band, continuum-subtracted image provides a panoramic view of Cas~A at subarcsecond resolution, tracing both shocked ejecta and shocked CSM. A comprehensive investigation of the 2013 \img\ was conducted by \citet{koo18}. Below, we summarize the main morphological features relevant to this work.

The most prominent structure is the bright, almost circular ring with a radius of $\sim$1\farcm7 (1.7~pc at 3.4~kpc), commonly referred to as the main ejecta shell. This ring is incomplete, with clear disruptions in the eastern and western sectors. It predominantly traces dense C- and O/Ne-burning ejecta material (e.g., O, Ne, Mg, Si, S) that has been heated by the reverse shock. The \fe\ emission is attributed to Fe$^+$ ions in the postshock cooling region. The origin of Fe in the shell may be a mixture of pre-SN Fe (inherited from the progenitor star) and newly synthesized Fe during the SN explosion \citep[see][]{koo18}. Previous 3D Doppler mapping studies have shown that the shocked ejecta in the main ejecta shell are largely organized into circular ring-like structures, possibly produced by post-explosion input of energy from plumes of radioactive $^{56}$Ni-rich ejecta \citep{delaney10,milisavljevic13}. These ring structures lie on the surface of a sphere that presumably traces the reverse shock front, so a spherical shell model could be feasible. Note that the \img\ is a continuum-subtracted narrow-band image covering velocities of $\pm$2600\kms, so ejecta with LOS velocities beyond this range can appear as negative features (e.g., interior filaments including the southern portion of the crown-like ring structure in the northern interior). Despite this limitation, the velocity coverage encompasses the majority of the ejecta in the main shell, as illustrated in Figure~9 of \citet{delaney10} and Figure~4 of \citet{milisavljevic13}.

Outside the main shell, both inside and outside the SNR, numerous compact knots are visible, classified as QSFs or FMKs (see Section~\ref{sec:catalog}). In the figure, QSFs are outlined in magenta, and FMKs in cyan. The brightest knots are mostly QSFs. Among the QSFs, the most striking feature is the large arc in the southern region, long recognized in optical studies \citep{lawrence95}. A dense population of QSFs is also apparent in the west. Due to heavy extinction, this population was mostly invisible in optical studies and was first revealed in the 2013 \img\ \citep{koo18}.

Interior to the main shell, the image reveals faint, diffuse emission with filamentary or clumpy morphology. This diffuse component is attributed to unshocked SN ejecta, most likely dominated by \si\ 1.645~\micron\ emission, rather than \fe. This interpretation is supported by their strong spatial correlation with Spitzer [\ion{Si}{2}] 34.8~\micron\ emission and the absence of interior \fe\ MIR lines in these regions \citep{koo18}. The emission likely arises from cold ($T \lesssim$ 100~K), weakly ionized Si that has been photoionized by UV and X-ray photons from reverse shock \citep{raymond18}. This diffuse component is too faint for reliable kinematic studies and is therefore not analyzed in this work.

\subsection{2020-2013 Difference Image \label{sec:difference}}

The lower frame of Figure~\ref{fig:overview} shows the difference image obtained by subtracting the 2013 \img\ from the 2020 one. In this image, regions that have brightened over the seven-year interval appear in white, while those that have faded appear in black. FMKs located beyond the main ejecta shell manifest as white-black pairs that are radially displaced by more than a few arcseconds, reflecting their high velocities.

Along the main shell, the outer rim generally appears white and the inner rim black, reflecting the outward motion of shocked ejecta. Particularly bright filaments and clumps with adjacent white-black contrast are visible in the northern and southwestern shell regions. This morphological pattern is consistent with dense ejecta expanding at typical speeds of $\sim$ 5000\kms, corresponding to a proper motion of $\sim$ 2\arcsec\ (or 0.036~pc) over seven years. At this spatial scale, brightening and fading features often appear contiguous, producing characteristic white-black pairs. This morphology also implies that the reverse shock is expanding outward in the observer's frame. If the reverse shock were instead moving inward in the plane of the sky, one would expect to see a brightened inner rim, as newly shocked ejecta begin to radiate. The absence of such inner brightening across most of the shell suggests that the reverse shock generally propagates outward in the sky, consistent with previous optical studies \citep[e.g.,][]{fesen25}. However, in certain regions, the morphology is complex, particularly in the eastern and western disrupted portions of the shell and in the southern part. In these areas, the expected white-black associations are less obvious or ambiguous. We defer a detailed discussion of these regions to Section~\ref{sec:shell}.

QSFs, due to their low velocities, remain essentially fixed in position over this time span and are therefore largely absent from the difference image. However, QSFs that experienced significant brightness changes appear as white or black features, depending on whether they brightened or faded between epochs. In the southern region outside the main ejecta shell, many QSFs appear white, indicating an overall increase in brightness. In contrast, QSFs in the southeastern interior of the SNR tend to appear black, suggesting they have dimmed. In the disrupted western region, some QSFs display a dark core encircled by a faint white halo, possibly reflecting surface brightness variations within the knots or observational seeing.

\section{Revised Catalog of Compact Knots \label{sec:catalog}}

\subsection{Identification of Knots \label{ssec:identify}}

We extracted knots following the same procedure described in Section~4 of \citet{koo18} to ensure consistency in comparison. First, we smoothed the continuum-subtracted image using the IDL SMOOTH procedure with a three-by-three-pixel boxcar. Next, we drew contours on the smoothed image and identified knots, excluding stars, artificial patterns, and contours smaller than five pixels. The background rms noise in the smoothed 2020 \img\ is 1.18 $\times 10^{-18}$ \unit\ pixel$^{-1}$. We applied the same threshold of $4 \times 10^{-18}$ \unit\ pixel$^{-1}$\ as used in \citet{koo18}. Due to the low sensitivity of the 2020 \img, this threshold corresponds to 3.4$\sigma$, compared to the 5$\sigma$ threshold in \citet{koo18}. This threshold remains low enough to capture faint emissions while effectively excluding background noise, making it suitable for comparison with earlier results, such as flux variation. Some knots in close proximity appeared merged at this threshold. Three such groups were identified (Knots 6--9, 23--24, and 168--169). Although connected, their individual shapes were distinguishable, and we therefore applied a higher threshold to catalog them separately. Conversely, several knots in Table~1 of \citet{koo18} (hereafter 2013 Catalog) appeared fragmented in the 2020 \img; in these cases, each fragment was labeled by a lowercase letter, and their parameters are tabulated individually. For the analysis presented in this paper, fragments with the same parent number are treated collectively as a single knot. In total, we identified and cataloged 203 knots outside the main ejecta shell.

In the process of catalog construction, we also identified regularly spaced artifacts surrounding bright stars, likely introduced during data reduction. Some of these features were included in the 2013 Catalog, as they were not distinguishable in the 2013 data alone. However, a comparison with the 2020 \img\ revealed that the movement of these features is inconsistent with genuine ejecta, and they do not appear in the 2022 \jwst\ F162M image \citep{milisavljevic24}, unlike other genuine knots. Consequently, we have excluded four features from our current catalog (Knots 37, 101, 143, and 156 from the 2013 Catalog).

In addition to the knots outside the main shell, we also cataloged compact features located on the main ejecta shell. As noted by \cite{koo18}, defining a ``knot'' on the ejecta shell is subjective, and cataloging all those knot-like features may not be meaningful; therefore, we limit the identification to bright features. We first identified QSFs using the same thresholds employed in \citet{koo18}. Next, we extracted several newly visible QSFs by drawing contours that isolated them from surrounding ejecta components. These QSFs were visible in the 2013 \img\ with threshold values ranging from $6.4 \times 10^{-18}$ to $1.1 \times 10^{-17}$ \unit\ pixel$^{-1}$, although they were not previously cataloged. These threshold values were applied to the 2020 \img\ whenever possible, with adjustments made if nearby features were connected to the contour or if no contours appeared at the original threshold. We identified 57 QSFs on the shell, including 51 from the 2013 Catalog. Their parameters were derived in the same manner as those of the knots outside the shell. Additionally, we included three bright FMKs that did not meet the criteria outlined above. These knots are located just outside the main shell and are prominent in the 2020 \img. We extracted them by applying a higher threshold.

\startlongtable
\begin{deluxetable*}{lllrrrrcccrc}
\tabletypesize{\scriptsize}
\setlength{\tabcolsep}{3pt}
\tablecaption{A Catalog of Knots in the 2020 \fe+\si\ Image\label{tab:catalog}}
\tablewidth{0pt}
\tablecolumns{12} \tablehead{
\colhead{ID} & \colhead{$\alpha$(J2000)} & \colhead{$\delta$(J2000)} &
\colhead{$D_{\rm max}$} & \colhead{$D_{\rm min}$} & \colhead{$\psi_{\rm ellipse}$} &
\colhead{Area} & \colhead{$F_{\rm obs}$} & \colhead{$F_{\rm ext-corr}$} &
\colhead{$\Delta r$} & \colhead{$\theta_{\Delta r}$} & \colhead{Class} \\
\colhead{ } & \colhead{ } & \colhead{ } & \colhead{(arcsec)} &
\colhead{(arcsec)} & \colhead{(deg)} & \colhead{(arcsec$^2$)} &
\colhead{(\unit)} & \colhead{(\unit)} &
\colhead{(arcsec)} & \colhead{(deg)} & \colhead{ }}
\decimalcolnumbers
\startdata
1$^{S}$ & 23:23:25.362 & 58:47:07.47 & 2.33 & 2.07 & 128 & 3.77 & 1.21E-14 & 6.67E-14 & 0.13 & -173.0 & QSF\\
2$^{S}$ & 23:23:25.165 & 58:47:05.42 & 1.54 & 1.00 & 57 & 1.22 & 2.40E-15 & 1.38E-14 & 0.06 & -173.8 & QSF\\
3 & 23:23:24.196 & 58:46:31.07 & 1.48 & 0.93 & 126 & 1.01 & 1.60E-16 & 9.80E-16 & 0.42 & -66.4 & QSF$^{a}$\\
4 & 23:23:24.115 & 58:46:47.71 & 1.92 & 1.01 & 20 & 1.38 & 2.24E-16 & 1.21E-15 & 0.40 & 34.5 & QSF$^{a}$\\
5 & 23:23:23.699 & 58:46:38.27 & 1.52 & 1.21 & 92 & 1.42 & 3.49E-16 & 1.77E-15 & 0.10 & -159.8 & QSF\\
6$^{M}$ & 23:23:23.040 & 58:46:34.80 & 3.48 & 1.97 & 78 & 4.90 & 2.63E-15 & 1.27E-14 & 0.11 & -103.5 & QSF\\
7$^{M}$ & 23:23:23.565 & 58:46:32.33 & 4.85 & 4.54 & 133 & 16.53 & 1.64E-14 & 7.96E-14 & 0.06 & -38.6 & QSF\\
8$^{M}$ & 23:23:22.567 & 58:46:33.26 & 4.30 & 1.39 & 96 & 4.46 & 1.93E-15 & 1.06E-14 & 0.08 & 87.2 & QSF\\
9$^{M}$ & 23:23:22.804 & 58:46:30.73 & 1.80 & 1.20 & 73 & 1.70 & 7.76E-16 & 3.76E-15 & 0.02 & 162.9 & QSF$^{a}$\\
10 & 23:23:22.124 & 58:46:26.73 & 7.36 & 3.86 & 85 & 20.54 & 8.27E-15 & 4.55E-14 & 0.41 & 85.4 & QSF\\
\enddata
\tablecomments{(1) Knot identification number; (2)--(3) central coordinates from ellipse fitting;
(4)--(5) major and minor axes from ellipse fitting;
(6) P.A. of the major axis measured from north to east;
(7) area enclosed by the contour; (8) observed flux; (9) extinction-corrected flux;
(10) angular displacement between 2013 and 2020, measured using the intensity-weighted centroid; (11) P.A. of the displacement vector, measured from north to east;
(12) classification of each knot.}
\tablenotetext{*}{Central coordinates from the 2013 Catalog.}
\tablenotetext{\it a}{Classified as cQSFs in 2013.}
\tablenotetext{\it M}{Knots merged in the 2020 image. A higher threshold is applied for these knots.}
\tablenotetext{\it S}{Knots on shell. A higher threshold is applied for these knots.}
\tablenotetext{}{(This table is available in its entirety in machine-readable form.)}
\end{deluxetable*}

Table~\ref{tab:catalog} (hereafter 2020 Catalog) summarizes the parameters of 263 knots in total, comprising 203 located outside the main ejecta shell and 60 on the shell itself. Knots previously cataloged in 2013 retain their original IDs, while those newly identified in the 2020 image are assigned IDs 310--348. The geometrical parameters --- specifically, the central coordinates, major and minor axes, and position angles (P.A.; from north to east) --- were obtained by fitting each contour by an ellipse using the FIT\_ELLIPSE procedure from the Coyote IDL Library.\footnote{The \texttt{Fit\_Ellipse} routine is part of the Coyote IDL Program Libraries (\url{https://github.com/davidwfanning/idl-coyote}).} The area and flux were directly calculated from the contours. The table also includes extinction-corrected fluxes. We applied extinction corrections to the observed fluxes following the method of \citet{koo18}. Specifically, for most knots, we used the column density map from \citet{hwang12}, while for those located outside this map, we used $Herschel$ SPIRE 250~\micron\ image. To convert column density ($N_{\rm H}$) to visual extinction ($A_V$), we adopted a ratio of $A_V/N_{\rm H}=1.87\ \times\ 10^{21}$ cm$^{-2}$ mag$^{-1}$ and used $A$(1.644~\micron)$/A_V$ = 1/5.4. As noted by \citet{koo18}, the column density map of \citet{hwang12}, derived from X-ray spectral analysis, may overestimate extinction for heavily obscured knots, leading to an overestimation of extinction-corrected fluxes of up to $\lesssim$ 60\%.

\subsection{Classification of Knots \label{ssec:classify}}

The knots in the 2020 Catalog were matched with those in the 2013 Catalog based on their center positions. Initially, knots were paired within a separation of 1\farcs24, which corresponds to the maximum proper motion of QSFs (0\farcs18 yr$^{-1}$) reported by \citet{vandenbergh85}. For FMK matching, the knot center positions of the 2013 catalog were shifted up to 2\% from the expansion center. This threshold was determined based on the known age of the SNR \citep[$\sim$350~years;][]{thorstensen01} and the time elapsed between the two epochs (6.91~years). Following the initial pairing, a visual inspection of the shapes and positions of the knots was conducted to validate the matches. This inspection involved comparing the 2020 \img\ with the 2013 \img, aligning both to the sky frame and free-expanding ejecta frame. In total, 224 of the 309 knots cataloged by \citet{koo18} were matched, including cases where individual knots had fragmented or merged. The remaining 85 knots from the 2013 Catalog, predominantly FMKs, have no clear counterparts in the 2020 data, while 39 additional knots appear only in the 2020 Catalog. 

We classified the knots common to both the 2013 and 2020 Catalogs according to their measured positional shifts. Knots displaced by more than 0\farcs76 were categorized as FMKs, while those with smaller shifts were classified as QSFs. The threshold value of 0\farcs76 corresponds to the largest proper motion (0\farcs11 yr$^{-1}$) observed among the QSFs in \citet{koo18}. As noted in the paper, 0\farcs11 yr$^{-1}$, which corresponds to 1800\kms, is unusually high for QSFs and was likely caused by a change in the brightness distribution rather than the actual motion. In our analysis as well, contour variability due to brightness changes significantly affected the measured proper motions. According to \citet{koo25}, who measured proper motions of optically bright QSFs using \hst\ images spanning 15~years, the largest proper motion was about 0\farcs03 yr$^{-1}$ ($\approx$500\kms). For this reason, the positional shifts derived in this section are used only to distinguish QSFs from FMKs.

For the classification of the knots without counterparts in the 2013 Catalog (Knots 310--348), we searched for faint features in the 2013 \img\ that had not been previously assigned to any knots. Most knots outside the main shell had faint counterparts in the 2013 \img\ below the original detection threshold and could thus be classified. Three knots (313, 316, and 326) within the main shell had no visible counterparts in the 2013 \img. To identify them, we compared the 2020 image to the \jwst\ NIRCam F162M image obtained in November 2022. All three were classified as QSFs, as their counterparts showed negligible proper motion.

The classification results are summarized in the last column of Table~\ref{tab:catalog}. In total, we identify 132 QSFs and 131 FMKs. Among the newly cataloged knots (IDs 310--348), 17 are classified as QSFs and 22 as FMKs. Upon visual inspection, we found four knots (19, 25, 61, and 95) that had been misclassified as FMKs; their apparently large motions were instead caused by overlapping ejecta (Knot 61) or by brightness variations of sub-clumps (Knots 19, 25, and 95). These knots have been reclassified as QSFs.

There are several caveats regarding the knot classification in Table~\ref{tab:catalog}. As noted above, the threshold positional shift separating FMKs and QSFs (0\farcs76) is significantly larger than the shift expected from typical QSF proper motions ($\lesssim$0\farcs03~yr$^{-1}$; \citealt{koo25}). Consequently, QSFs with measured shifts exceeding $\sim$0\farcs21, corresponding to the expected displacement over seven years for a proper motion of 0\farcs03~yr$^{-1}$, should be treated with caution. In addition, the classification is based solely on projected motion in the plane of the sky and does not account for the LOS velocity component. In principle, FMKs moving predominantly along the LOS in the central regions could therefore be misclassified as QSFs. However, all knots classified as QSFs lie at projected radii greater than 25\farcs, and those between 25\farcs\ and 35\farcs\ generally exhibit positional shifts well below those expected for FMKs. The only exceptions are Knots~110 and~111, which are faint and likely affected by noise-induced uncertainties in their measured shifts. We therefore conclude that contamination of the QSF sample by FMKs with predominantly LOS motion is unlikely.

\subsection{Changes in the Knot Catalog since 2013 \label{ssec:variation}}

\subsubsection{Knot Population \label{sssec:population}}

A total of 263 knots have been identified in the 2020 \img, of which 132 are classified as QSFs and 131 as FMKs. In comparison, the previous study by \citet{koo18} reported 309 knots, including 130 QSFs (36 of which are candidates) and 179 FMKs (12 of which are candidates). The candidates refer to knots for which proper motion could not be measured. \citet{koo18} classified the knots in the 2013 \img\ by measuring their proper motions relative to a Palomar 2008 \fe$+$\si\ image, which had lower sensitivity and resolution. The majority of knots lacked counterparts in the 2008 image; therefore, the authors also used the \hst\ F098M and F850LP images obtained in 2011 and 2004 for classification. Knots without counterparts in the \hst\ images were classified as FMK candidates if they were located around the NE or the western jet area, or as QSF candidates if they were found in regions predominantly occupied by QSFs or had counterparts in H$\alpha+$[\ion{N}{2}] $\lambda\lambda$6548, 6583 images. In the present work, the classifications of these candidate sources have been finalized.

\begin{deluxetable}{lcc}
\tablecaption{Summary of Changes in the Knot Population from the 2013 Catalog \label{tab:knotcount}}
\tablehead{ \colhead{} & \colhead{Number of Knots} & \colhead{} \\
\colhead{} & \colhead{in the 2020 Catalog} & \colhead{Details} }
\startdata
QSF & 132 & (92 QSF + 20 cQSF + 3 cFMK) in the 2013 catalog + 17 newly identified \\
FMK & 131 & (106 FMK + 1 cFMK + 2 cQSF) in the 2013 catalog + 22 newly identified \\
\hline
Total & 263 & \\
\enddata
\tablecomments{Number of disappeared knots in the 2013 catalog: 2 QSF + 14 cQSF, 61 FMK + 8 cFMK (cQSF = QSF candidate, cFMK = FMK candidate)}
\tablecomments{Disappeared QSFs = Knots 93, 96; Disappeared cQSFs = Knots 14, 16, 17, 37, 91, 98, 99, 101, 116, 117, 228, 266, 307, 308}
\end{deluxetable}

Table~\ref{tab:knotcount} summarizes the changes in the numbers of QSFs and FMKs, compared to the previous catalog. Firstly, two QSF candidates from the earlier study were reclassified as FMKs, while three FMK candidates were reclassified as QSFs. Note that the candidates in \citet{koo18} refer to knots for which proper motion could not be measured. Secondly, 16 QSFs (including 14 candidates) and 69 FMKs (including 8 candidates) were no longer present in the 2020 \img. Most of the QSFs that have disappeared, with the exception of two, are faint knots that were categorized as candidates. The other two cases involve bright QSFs located on the main ejecta shell, corresponding to the previous knot numbers 135 and 159. Neither of these knots had an optical counterpart in \citet{koo18}, and neither is present in the 2020 \img. It is likely that they were mistaken for features within the ejecta shell. The classification of the remaining knots remained consistent, confirming 20 QSF candidates and an FMK candidate. Additionally, 17 new QSFs and 22 new FMKs were detected in the 2020 \img.

\begin{deluxetable}{ l c c } 
\tabletypesize{\scriptsize}
\tablecaption{\fe\ 1.644~\micron\ Properties of QSFs and FMKs \label{tab:property}}
\tablehead{\colhead{Parameter} & \colhead{QSF} & \colhead{FMK}} 
\startdata
Number of identified knots & 132 & 131 \\
Geometrical radius (arcsec) & 1.10 (0.25--4.70) & 0.68 (0.25--2.39) \\
$D_{\rm min}/D_{\rm max}$ & 0.67 (0.23--1.00) & 0.71 (0.16--1.00) \\
\underline{Observed} & & \\
Flux ($10^{-15}$ \unit) & 1.67 (0.042--236) & 0.31 (0.042--12.4)\\
Brightness ($10^{-16}$ \unit & 4.04 (1.31--212) & 2.01 (1.39--37.2)\\
\hspace{3mm} arcsec$^{-2}$) & & \\
Total flux ($10^{-12}$ \unit) & 1.29 & 0.13\\
\underline{Extinction corrected} & & \\
Flux ($10^{-15}$ \unit) & 9.83 (0.24--1114) & 1.21 (0.14--52.2)\\
Brightness ($10^{-16}$ \unit & 22.1 (5.75--1025) & 8.77 (4.86--122)\\
\hspace{3mm} arcsec$^{-2}$) & & \\
Total flux ($10^{-12}$ \unit) & 7.64 & 0.54\\
Luminosity ($L_\sun$) & 2.76 & 0.20 \\
\enddata
\tablecomments{The quoted values are the median, and the numbers in parentheses are the minimum and maximum.}
\end{deluxetable}

\subsubsection{Flux Variability \label{sssec:variability}}

Table~\ref{tab:property} shows the properties of QSFs and FMKs. For the QSFs, the total observed and extinction-corrected fluxes in 2020 are $1.29\times10^{-12}$ and $7.64\times10^{-12}$ \unit, respectively, nearly identical to the 2013 values. The total observed and extinction-corrected fluxes of the FMKs are $0.13\times10^{-12}$ and $0.54\times10^{-12}$ \unit, respectively, which are slightly higher than those measured in 2013. Although their total fluxes changed little over seven years, individual knots show noticeable brightness variations. In the following, we present the flux variability of QSFs and FMKs separately and describe the associated error analysis.

\begin{figure}
\fig{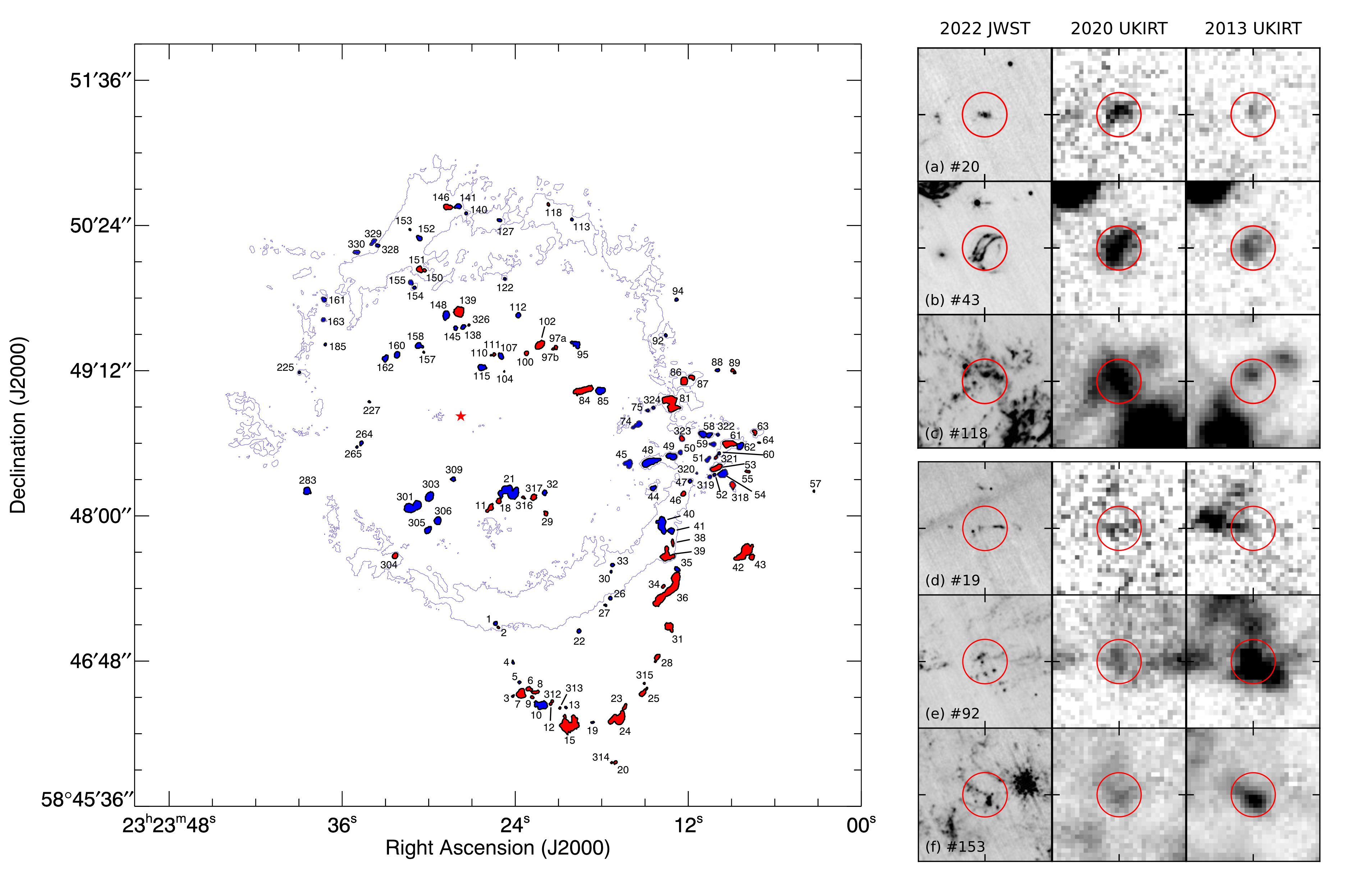}{0.9\textwidth}{} 
\caption{Left: A revised finding chart of QSFs. QSFs colored blue indicate darkening in 2020, and red indicates brightening. The blue contour shows the main ejecta shell in 2020 \img. The red star represents the explosion center. Right: Example of QSFs with significant flux change. From left to right, the images correspond to the 2022 \jwst\ image, the 2020 UKIRT image, and the 2013 UKIRT image. Knot numbers are shown on the left, and the red circle has a radius of 1\arcsec. \label{fig:finding}}
\end{figure}

\begin{figure}
\fig{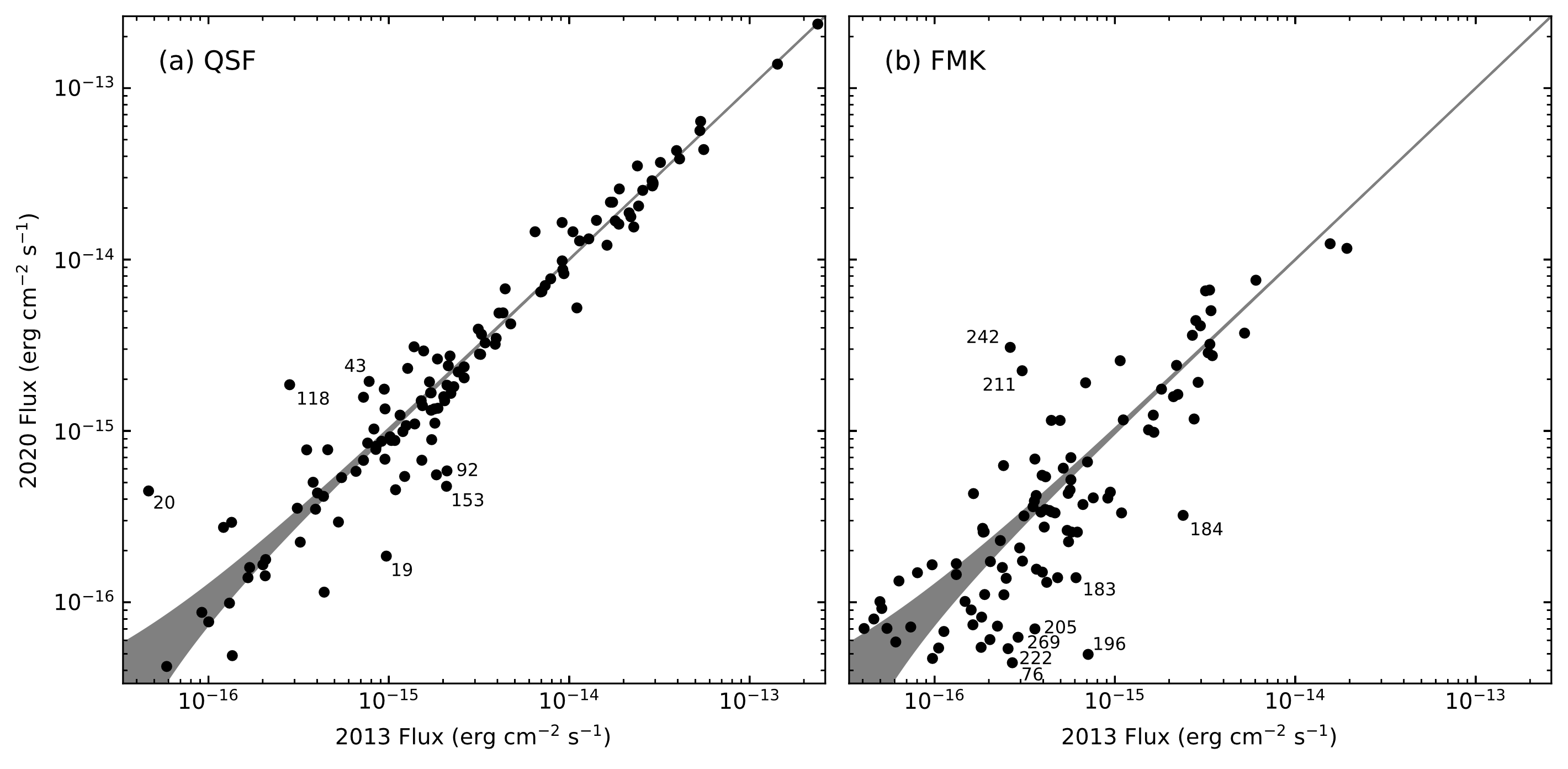}{0.9\textwidth}{} 
\caption{(a) A comparison plot of QSF flux between 2013 and 2020. (b) The same as (a) but for FMKs. The gray area indicates the 1$\sigma$ confidence band of unchanged synthetic knots for reference. Knot numbers with flux variations exceeding a factor of 4 are labeled. \label{fig:fluxcomp}}
\end{figure}

\paragraph{QSFs. \label{par:qsf}}

Figure~\ref{fig:finding} presents a finding chart for the 132 QSFs listed in the 2020 Catalog. In the figure, QSFs that brightened in 2020 are marked in red, while those that dimmed are shown in blue. The knots along the southern arc structure generally brightened, while the knots in the interior and western regions generally darkened. The brightness variations of the individual knots are shown in Figure~\ref{fig:fluxcomp}(a). In total, 53 QSFs increased in brightness, while 79 decreased. Although the total flux of QSFs shows a difference of less than 2\%, individual knots show significant flux variations, in some cases by an order of magnitude. Knots with pronounced flux changes are highlighted on the right side of Figure~\ref{fig:finding}: the top panels (a)--(c) show QSFs that brightened in 2020, while the bottom panels (d)--(f) display QSFs that dimmed in 2020. Their corresponding \jwst\ F162M images from 2022 are also shown.

The brightening of the QSF knots along the southern arc structure is particularly noteworthy. This arc is a prominent feature in optical images and has been recognized in previous studies \citep{lawrence95}. The observed brightening likely indicates that shocks are currently propagating into these dense clumps, since QSFs generally fade optically once the entire clump has been shocked \citep[e.g.,][]{koo25}. Indeed, for one of the QSFs at the southern end of the arc (QSF~24), \citet{koo20} found evidence that the knot is partially destroyed, with some portions still unshocked. For a typical QSF of radius $a$, the characteristic lifetime can be estimated by the cloud-crushing timescale, $\tau_{cc}\equiv a/v_{cs}=100(a/0.01~{\rm pc})(v_{cs}/100$\kms)~years, where $v_{cs}$ is the velocity of the shock propagating into the clump \citep{klein94}. After approximately $\tau_{cc}$, the knot is expected to be disrupted by interactions with the SN blast wave and become optically faint. Observationally, most QSFs in the southern arc (Figure~\ref{fig:finding}) are visible in early optical images \citep{vandenbergh85,koo18,koo25}, suggesting that their current ages exceed 60--70~yr, but remain shorter than the cloud-crushing timescale. This is consistent with their ongoing brightening, suggesting they are in an intermediate evolutionary stage --- currently being overtaken by the shock but not yet fully disrupted. 

\begin{figure}
\fig{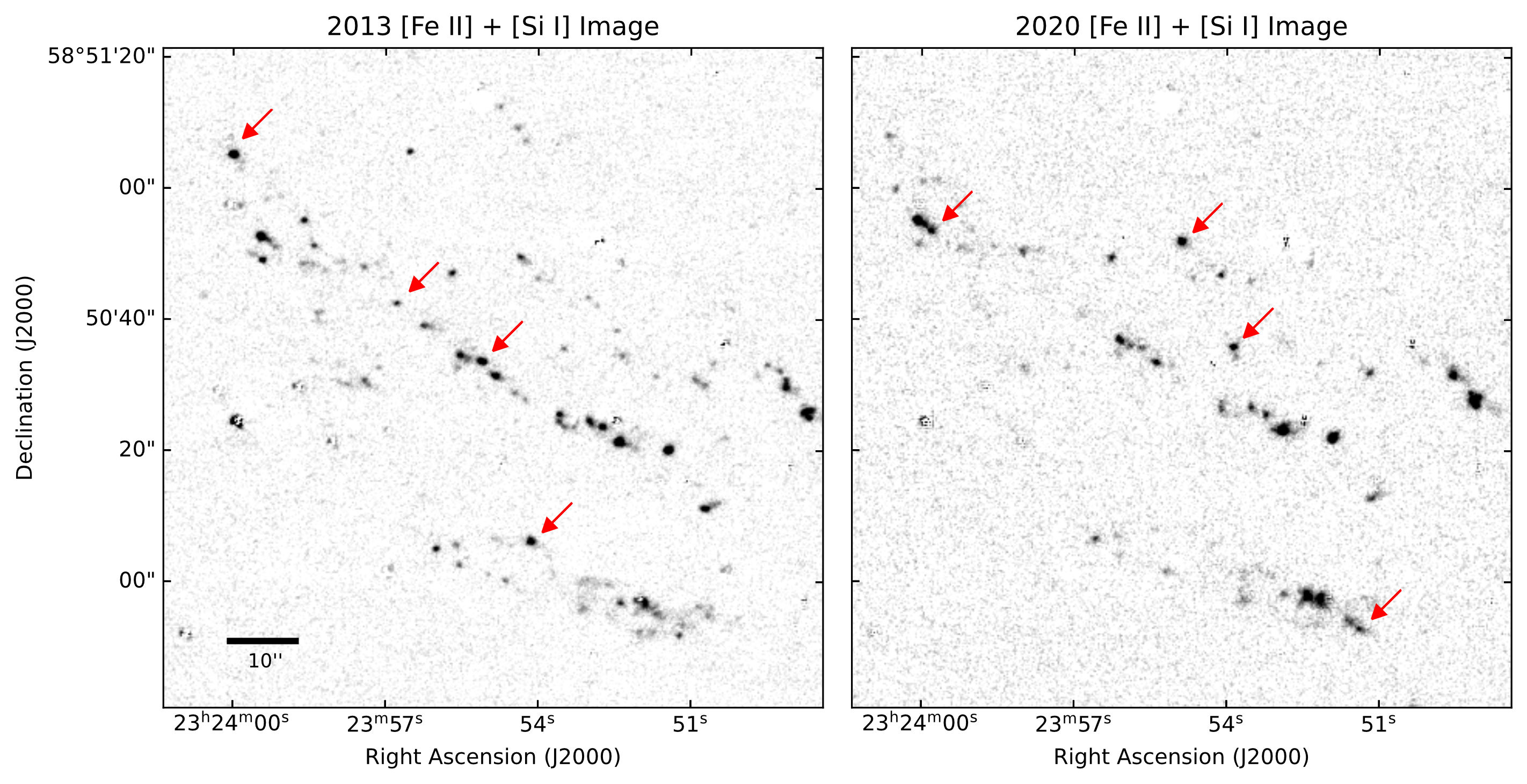}{0.9\textwidth}{} 
\caption{Magnified image of the NE~jet region. On the left is the 2013 image, while on the right is the 2020 image. Both images are displayed in grayscale with an intensity range from $0$ to $2 \times 10^{-17}$ \unit\ pixel$^{-1}$ , and red arrows mark knots that appear bright in only one of the two epochs. \label{fig:fmkchange}}
\end{figure}

\paragraph{FMKs. \label{par:fmk}} 

Figure~\ref{fig:fluxcomp}(b) shows flux variation of FMKs between 2013 and 2020. Note that FMKs are generally fainter than QSFs. The majority of FMKs have undergone flux changes by a factor of a few over 7 years, whereas QSFs show little flux variation, except for some outliers. We define the logarithmic flux ratio as $\log_2(F_{2020}/F_{2013})$, which quantifies the relative change in brightness. Mean and standard deviation of the ratio are $-0.05\pm0.75$ for QSFs and $-0.30\pm1.13$ for FMKs. These results indicate that the flux variation in FMK is greater than in QSF. The larger flux variability of FMKs may be associated with their metal-rich composition, which enables rapid radiative cooling \citep[e.g.,][]{raymond18a}, and may also be influenced by the complex structure of the ejecta knots or by interactions with a highly inhomogeneous CSM \citep[see][]{fesen11}.

A representative region where significant flux variations in FMKs are observed is the NE~Jet area, as shown in Figure~\ref{fig:fmkchange}. In the figure, the FMKs that have darkened and brightened considerably are marked in the 2013 and 2020 images, respectively. \citet{fesen11} reported significant flux variations in knots located in the NE~Jet over a short period of just nine months. The selected knots in that study exhibited a flux variation of $\log_2(F_{\mathrm Dec}/F_{\mathrm Mar}) = 1.07 \pm 1.44$ in the F850LP band \citep[Table~1 of ][]{fesen11}, which is considerably larger than the flux variations observed for the FMKs in our study. This discrepancy likely arises because the study of \citet{fesen11} specifically selected knots with large flux variations, whereas our study includes all knots without pre-selection. Additionally, faint knots with large flux variations may not have been extracted in our analysis if their brightness fell below the detection threshold, which could have contributed to the absence of such extreme variations in our results.

\paragraph{Error Analysis. \label{par:error}} 

In Figure~\ref{fig:fluxcomp}, the gray region indicates uncertainty in the derived fluxes. We generated synthetic images and performed Monte Carlo simulations to estimate the uncertainties and systematic errors associated with the current flux extraction method. Two-dimensional Gaussian knots were randomly placed across the image, and Gaussian noise was added based on the observed image sensitivity ($3.2\times10^{-18}$ \unit\ pixel$^{-1}$). The knot width ($\sigma$) was set to $0\farcs4$, consistent with that derived from fitting the observed data. Knot fluxes varied from $2.8 \times 10^{-16}$ to $1.0 \times 10^{-12}$ \unit\ in logarithmically evenly spaced over 72 steps. We generated 1000 images for each flux level and extracted fluxes following the identical procedure applied to the observed data. The gray region in Figure~\ref{fig:fluxcomp} indicates 1$\sigma$ uncertainty. These uncertainties are smaller than those of observational flux changes.

However, our flux measurement method excludes pixels below the threshold, so it makes the knots appear dimmer than they are. In our Monte Carlo simulation, the knot flux with $10^{-15}$ \unit\ loses about 20\%\ of its original flux, and the knot flux with $10^{-16}$ \unit\ is not detected with the current method. Since fainter knots lose a larger portion of the flux, this makes the ratio of 2013 and 2020 flux more deviated from unity. This simulation assumed each knot as a single Gaussian function, but there are multiple peaks or complex structures in most of the knot contours, so flux loss would be higher than the estimation.

\section{Deceleration of the Outer FMKs \label{sec:decel}}

\subsection{Proper motion of FMKs \label{sssec:fmkpm}}

The displacements of the knots listed in the 2020 Catalog were measured from their intensity-weighted centers and used primarily to distinguish QSFs from FMKs. In this section, we perform more accurate proper-motion measurements for the FMKs using the chi2\_shift function from the image-registration Python package. This function uses cross-correlation and the discrete Fourier transform upsampling algorithm to determine the sub-pixel displacement and its error between two images by minimizing a $\chi^2$ metric \citep{guizar08}. We have verified its performance on synthetic Gaussian test images and confirm that it recovers known positional offsets with sub‐pixel accuracy.

For each knot, we extracted square patches from the 2013 and 2020 epoch images, centered on the ellipse fit center described in Section~\ref{ssec:identify}. The side length of each patch was set to the larger of the two ellipse major axes, ensuring that the full knot is enclosed at both epochs. To estimate the uncertainty for each measurement, we computed the sample standard deviation of the pixel values within a 3$\times$3 neighborhood around the central pixel and used it as the error for that pixel. A limitation of our approach is that it follows the motion of the image patch rather than the knot itself, potentially biasing the measured proper motion when the surrounding emission is strong. Because the uncertainty is estimated from the statistics of all pixels within the patch, the resulting error bars tend to be smaller than those one would obtain only by the knot emission. Both of these effects are mitigated when the knot is bright and the background is faint.

We tested the robustness of our proper motion measurements by increasing the patch radius in one pixel increments, up to five pixels. Most of the knots are shifted less than one pixel, but some knots are shifted more than two pixels. In such cases, where the knots were very faint, severely changed, or wrongly matched to neighboring features, we excluded knots that shifted by more than two pixels during the robustness test. Applying these criteria resulted in the removal of eight knots (ID: 169, 202, 206, 208, 211, 217, 238, and 269).

Table~\ref{tab:pm} summarizes our proper motion results. For each knot identified by the same IDs as in 2020 Catalog, we give the projected distance $r$ and position angle $\theta$ relative to the explosion center of \citet{thorstensen01}, the proper‐motion magnitude $\mu$ and its direction $\theta_{\mu}$, the expansion index $m$ ($r\propto t^m$), and the transverse velocity $v_{\mu}$ assuming a distance of 3.4 kpc. The expansion index is computed as 

\begin{equation}
 m = \frac{d \ln r}{d \ln t} = \frac{\mu t}{r} \label{eqn:mvalue}
\end{equation}

where $t$(=342.4 years) is the time elapsed since the explosion, assumed to be the year 1671.3 \citep{thorstensen01}. The uncertainties on the proper motion vector combine the astrometric errors of each epoch with the measurement error from chi2\_shift. We extracted $\mu$, $\theta_\mu$, and $m$ using the proper motion vector, and calculated the corresponding error bars. For those eastern knots whose LOS velocities $v_{\rm LOS}$ were measured spectroscopically by \citet{koo23}, we also list $v_{\rm LOS}$ and the corresponding knot numbers from Table~3 of that work. When multiple components fall within a single knot aperture, we report their mean values and list all components. 

In Table~\ref{tab:pm}, several knots show expansion indices $m>1$. These knots often exhibit multiple brightness peaks, complicating the measurement of proper motion. High-resolution \hst\ observations have shown that many such knots are composed of smaller substructures \citep{fesen11}. Recent \jwst\ observations \citep{milisavljevic24} further reveal that more than half of the knots with $m>1$ exhibit multiple bright peaks, with different peaks sometimes dominating in brightness between the 2013 and 2020 epochs. These morphological changes can introduce systematic discrepancies in proper motion measurements when using lower resolution images.

\startlongtable
\begin{deluxetable}{cccccccccc}
\tablecaption{Proper motion of FMKs \label{tab:pm}}
\tablehead{
\colhead{ID} & \colhead{$r$} & \colhead{$\theta$} & \colhead{$\mu$} & \colhead{$\theta_\mu$} & \colhead{$m$} & \colhead{$v_\mu$} & \colhead{$v_{\rm LOS}$} & \colhead{K2023 No.} & \colhead{Class} \\ 
\colhead{} & \colhead{(arcsec)} & \colhead{(deg)} & \colhead{(arcsec\,yr$^{-1}$)} & \colhead{(deg)} & \colhead{} & \colhead{(\kms)} & \colhead{(\kms)} & \colhead{} & \colhead{}
}
\decimalcolnumbers
\startdata
56 & 189.15 & 261.5 & 0.54 $\pm$ 0.02 & 261.3 $\pm$ 2.9 & 0.971 $\pm$ 0.045 & 8658 $\pm$ 400 & $\cdots$ & $\cdots$ & SW~jet \\
70 & 198.05 & 262.6 & 0.60 $\pm$ 0.02 & 263.9 $\pm$ 2.3 & 1.037 $\pm$ 0.041 & 9675 $\pm$ 382 & $\cdots$ & $\cdots$ & SW~jet \\
71 & 201.71 & 263.6 & 0.62 $\pm$ 0.03 & 263.4 $\pm$ 2.2 & 1.060 $\pm$ 0.043 & 10072 $\pm$ 406 & $\cdots$ & $\cdots$ & SW~jet \\
73 & 215.32 & 262.5 & 0.62 $\pm$ 0.02 & 262.5 $\pm$ 2.2 & 0.984 $\pm$ 0.038 & 9987 $\pm$ 382 & $\cdots$ & $\cdots$ & SW~jet \\
76 & 195.30 & 268.5 & 0.55 $\pm$ 0.02 & 273.8 $\pm$ 2.5 & 0.957 $\pm$ 0.042 & 8809 $\pm$ 386 & $\cdots$ & $\cdots$ & SW~jet \\
77 & 197.06 & 270.6 & 0.57 $\pm$ 0.02 & 272.2 $\pm$ 2.4 & 0.997 $\pm$ 0.041 & 9261 $\pm$ 385 & $\cdots$ & $\cdots$ & SW~jet \\
80 & 211.35 & 272.0 & 0.70 $\pm$ 0.02 & 272.5 $\pm$ 2.1 & 1.132 $\pm$ 0.039 & 11269 $\pm$ 387 & $\cdots$ & $\cdots$ & SW~jet \\
82 & 178.96 & 276.0 & 0.48 $\pm$ 0.02 & 271.8 $\pm$ 2.9 & 0.924 $\pm$ 0.047 & 7789 $\pm$ 398 & $\cdots$ & $\cdots$ & SW~jet \\
83 & 190.90 & 274.7 & 0.55 $\pm$ 0.02 & 272.5 $\pm$ 2.5 & 0.984 $\pm$ 0.043 & 8854 $\pm$ 382 & $\cdots$ & $\cdots$ & SW~jet \\
93 & 95.84 & 295.5 & 0.22 $\pm$ 0.02 & 293.6 $\pm$ 6.3 & 0.771 $\pm$ 0.085 & 3480 $\pm$ 382 & $\cdots$ & $\cdots$ & SW~jet \\
96 & 96.58 & 297.9 & 0.27 $\pm$ 0.02 & 307.4 $\pm$ 5.1 & 0.954 $\pm$ 0.084 & 4343 $\pm$ 383 & $\cdots$ & $\cdots$ & SW~jet \\
106 & 204.61 & 322.0 & 0.58 $\pm$ 0.02 & 322.0 $\pm$ 2.4 & 0.975 $\pm$ 0.040 & 9403 $\pm$ 386 & $\cdots$ & $\cdots$ & Northern Area \\
109 & 139.06 & 324.4 & 0.39 $\pm$ 0.02 & 324.0 $\pm$ 3.5 & 0.959 $\pm$ 0.058 & 6284 $\pm$ 381 & $\cdots$ & $\cdots$ & Northern Area \\
124 & 120.50 & 344.7 & 0.33 $\pm$ 0.02 & 338.4 $\pm$ 4.2 & 0.929 $\pm$ 0.068 & 5273 $\pm$ 388 & $\cdots$ & $\cdots$ & Northern Area \\
125 & 139.65 & 347.3 & 0.35 $\pm$ 0.02 & 333.6 $\pm$ 3.9 & 0.850 $\pm$ 0.058 & 5593 $\pm$ 383 & $\cdots$ & $\cdots$ & Northern Area \\
133 & 140.26 & 348.7 & 0.39 $\pm$ 0.02 & 347.6 $\pm$ 3.5 & 0.946 $\pm$ 0.059 & 6255 $\pm$ 388 & $\cdots$ & $\cdots$ & Northern Area \\
136 & 123.63 & 354.2 & 0.35 $\pm$ 0.02 & 351.4 $\pm$ 3.9 & 0.969 $\pm$ 0.066 & 5646 $\pm$ 385 & $\cdots$ & $\cdots$ & Northern Area \\
137 & 135.22 & 356.9 & 0.41 $\pm$ 0.02 & 6.7 $\pm$ 3.3 & 1.044 $\pm$ 0.061 & 6652 $\pm$ 388 & $\cdots$ & $\cdots$ & Northern Area \\
142 & 133.09 & 357.8 & 0.49 $\pm$ 0.02 & 0.7 $\pm$ 2.8 & 1.253 $\pm$ 0.063 & 7856 $\pm$ 393 & $\cdots$ & $\cdots$ & Northern Area \\
173 & 205.56 & 60.5 & 0.56 $\pm$ 0.02 & 64.5 $\pm$ 2.4 & 0.940 $\pm$ 0.039 & 9105 $\pm$ 381 & $\cdots$ & $\cdots$ & NE~jet \\
180 & 238.36 & 59.8 & 0.70 $\pm$ 0.02 & 60.9 $\pm$ 2.0 & 1.006 $\pm$ 0.035 & 11301 $\pm$ 390 & +124 $\pm$ 1 & 5 & NE~jet \\
181 & 246.67 & 61.5 & 0.72 $\pm$ 0.02 & 62.6 $\pm$ 1.9 & 1.003 $\pm$ 0.034 & 11659 $\pm$ 396 & +428 $\pm$ 7 & 7 & NE~jet \\
183 & 270.31 & 62.3 & 0.80 $\pm$ 0.02 & 65.8 $\pm$ 1.7 & 1.007 $\pm$ 0.030 & 12824 $\pm$ 385 & $\cdots$ & $\cdots$ & NE~jet \\
184 & 284.37 & 61.5 & 0.84 $\pm$ 0.02 & 58.9 $\pm$ 1.6 & 1.005 $\pm$ 0.028 & 13472 $\pm$ 380 & +748 $\pm$ 9 & 8 & NE~jet \\
187 & 195.67 & 65.2 & 0.53 $\pm$ 0.02 & 67.6 $\pm$ 2.6 & 0.927 $\pm$ 0.042 & 8546 $\pm$ 383 & +284 $\pm$ 15 & 19, 20 & NE~jet \\
188 & 204.99 & 63.7 & 0.59 $\pm$ 0.02 & 63.8 $\pm$ 2.4 & 0.986 $\pm$ 0.041 & 9521 $\pm$ 398 & +429 $\pm$ 3 & 12 & NE~jet \\
189 & 211.71 & 64.3 & 0.63 $\pm$ 0.02 & 64.1 $\pm$ 2.1 & 1.020 $\pm$ 0.038 & 10179 $\pm$ 380 & +490 $\pm$ 2 & 17 & NE~jet \\
194 & 233.51 & 64.1 & 0.68 $\pm$ 0.02 & 63.9 $\pm$ 2.0 & 1.001 $\pm$ 0.035 & 11012 $\pm$ 387 & +483 $\pm$ 3 & 16 & NE~jet \\
195 & 238.24 & 64.0 & 0.73 $\pm$ 0.02 & 62.0 $\pm$ 1.9 & 1.044 $\pm$ 0.034 & 11716 $\pm$ 382 & +530 $\pm$ 2 & 13, 14 & NE~jet \\
197 & 251.22 & 66.3 & 0.74 $\pm$ 0.02 & 64.2 $\pm$ 1.9 & 1.001 $\pm$ 0.033 & 11852 $\pm$ 387 & -234 $\pm$ 4 & 23 & NE~jet \\
200 & 258.99 & 62.7 & 0.74 $\pm$ 0.02 & 61.7 $\pm$ 1.9 & 0.976 $\pm$ 0.032 & 11909 $\pm$ 390 & +832 $\pm$ 10 & 9 & NE~jet \\
204 & 266.87 & 63.5 & 0.78 $\pm$ 0.02 & 66.0 $\pm$ 1.7 & 1.002 $\pm$ 0.030 & 12599 $\pm$ 382 & $\cdots$ & $\cdots$ & NE~jet \\
207 & 274.21 & 63.4 & 0.75 $\pm$ 0.02 & 63.6 $\pm$ 1.8 & 0.933 $\pm$ 0.030 & 12059 $\pm$ 384 & +10 $\pm$ 2 & 11 & NE~jet \\
212 & 196.78 & 71.3 & 0.52 $\pm$ 0.02 & 66.4 $\pm$ 2.6 & 0.912 $\pm$ 0.042 & 8457 $\pm$ 384 & -2125 $\pm$ 2 & 31 & NE~jet \\
214 & 198.52 & 70.3 & 0.58 $\pm$ 0.02 & 69.8 $\pm$ 2.3 & 1.003 $\pm$ 0.041 & 9387 $\pm$ 384 & +709 $\pm$ 2 & 25, 26, 27 & NE~jet \\
215 & 202.69 & 70.5 & 0.58 $\pm$ 0.02 & 72.5 $\pm$ 2.4 & 0.976 $\pm$ 0.041 & 9319 $\pm$ 391 & $\cdots$ & $\cdots$ & NE~jet \\
218 & 207.83 & 71.3 & 0.67 $\pm$ 0.02 & 73.9 $\pm$ 2.1 & 1.098 $\pm$ 0.041 & 10754 $\pm$ 398 & $\cdots$ & $\cdots$ & NE~jet \\
219 & 208.85 & 70.2 & 0.69 $\pm$ 0.02 & 80.3 $\pm$ 2.0 & 1.122 $\pm$ 0.040 & 11045 $\pm$ 389 & $\cdots$ & $\cdots$ & NE~jet \\
221 & 220.36 & 71.2 & 0.65 $\pm$ 0.02 & 72.5 $\pm$ 2.1 & 1.001 $\pm$ 0.038 & 10395 $\pm$ 393 & $\cdots$ & $\cdots$ & NE~jet \\
224 & 231.90 & 70.9 & 0.68 $\pm$ 0.02 & 71.0 $\pm$ 2.0 & 1.010 $\pm$ 0.036 & 11034 $\pm$ 388 & -1199 $\pm$ 5 & 28 & NE~jet \\
226 & 179.21 & 73.8 & 0.53 $\pm$ 0.02 & 74.0 $\pm$ 2.6 & 1.004 $\pm$ 0.046 & 8476 $\pm$ 389 & -2017 $\pm$ 4 & 34 & NE~jet \\
231 & 157.39 & 84.9 & 0.44 $\pm$ 0.02 & 82.3 $\pm$ 3.1 & 0.958 $\pm$ 0.053 & 7102 $\pm$ 389 & $\cdots$ & $\cdots$ & Fe~K~plume \\
232 & 159.82 & 86.3 & 0.45 $\pm$ 0.03 & 86.6 $\pm$ 3.2 & 0.968 $\pm$ 0.054 & 7290 $\pm$ 405 & -1801 $\pm$ 4 & 42, 43 & Fe~K~plume \\
234 & 165.83 & 85.1 & 0.46 $\pm$ 0.02 & 86.1 $\pm$ 3.0 & 0.946 $\pm$ 0.050 & 7394 $\pm$ 393 & -775 $\pm$ 6 & 40, 41 & Fe~K~plume \\
235 & 167.21 & 84.7 & 0.48 $\pm$ 0.02 & 84.2 $\pm$ 2.9 & 0.988 $\pm$ 0.048 & 7787 $\pm$ 380 & -1286 $\pm$ 5 & 39 & Fe~K~plume \\
237 & 186.14 & 84.6 & 0.53 $\pm$ 0.02 & 85.0 $\pm$ 2.6 & 0.980 $\pm$ 0.045 & 8596 $\pm$ 391 & $\cdots$ & $\cdots$ & Fe~K~plume \\
239 & 158.00 & 88.6 & 0.44 $\pm$ 0.02 & 92.1 $\pm$ 3.1 & 0.945 $\pm$ 0.052 & 7038 $\pm$ 385 & -1162 $\pm$ 1 & 45, 47 & Fe~K~plume \\
241 & 162.01 & 88.0 & 0.44 $\pm$ 0.02 & 87.8 $\pm$ 3.1 & 0.938 $\pm$ 0.051 & 7161 $\pm$ 387 & -1189 $\pm$ 2 & 46 & Fe~K~plume \\
242 & 170.09 & 88.6 & 0.37 $\pm$ 0.02 & 83.7 $\pm$ 3.7 & 0.745 $\pm$ 0.048 & 5971 $\pm$ 386 & $\cdots$ & $\cdots$ & Fe~K~plume \\
244 & 136.64 & 97.0 & 0.38 $\pm$ 0.02 & 95.4 $\pm$ 3.6 & 0.947 $\pm$ 0.060 & 6098 $\pm$ 385 & +755 $\pm$ 2 & 50 & Fe~K~plume \\
246 & 156.63 & 94.7 & 0.44 $\pm$ 0.02 & 92.8 $\pm$ 3.2 & 0.957 $\pm$ 0.053 & 7061 $\pm$ 389 & $\cdots$ & $\cdots$ & Fe~K~plume \\
249 & 163.80 & 94.6 & 0.47 $\pm$ 0.02 & 93.6 $\pm$ 3.0 & 0.979 $\pm$ 0.051 & 7555 $\pm$ 394 & $\cdots$ & $\cdots$ & Fe~K~plume \\
250 & 165.52 & 97.2 & 0.50 $\pm$ 0.02 & 93.0 $\pm$ 2.7 & 1.027 $\pm$ 0.049 & 8007 $\pm$ 385 & +279 $\pm$ 1 & 51, 52 & Fe~K~plume \\
253 & 141.63 & 101.1 & 0.38 $\pm$ 0.02 & 99.3 $\pm$ 3.7 & 0.923 $\pm$ 0.060 & 6157 $\pm$ 397 & $\cdots$ & $\cdots$ & Fe~K~plume \\
255 & 145.84 & 100.9 & 0.41 $\pm$ 0.02 & 99.9 $\pm$ 3.3 & 0.972 $\pm$ 0.057 & 6682 $\pm$ 393 & $\cdots$ & $\cdots$ & Fe~K~plume \\
257 & 151.80 & 98.3 & 0.42 $\pm$ 0.02 & 97.2 $\pm$ 3.5 & 0.951 $\pm$ 0.054 & 6806 $\pm$ 388 & $\cdots$ & $\cdots$ & Fe~K~plume \\
259 & 159.17 & 101.4 & 0.48 $\pm$ 0.02 & 101.5 $\pm$ 2.8 & 1.038 $\pm$ 0.051 & 7785 $\pm$ 381 & +271 $\pm$ 2 & 53, 54 & Fe~K~plume \\
262 & 165.12 & 101.0 & 0.50 $\pm$ 0.02 & 104.5 $\pm$ 2.8 & 1.042 $\pm$ 0.050 & 8105 $\pm$ 386 & $\cdots$ & $\cdots$ & Fe~K~plume \\
267 & 102.96 & 106.6 & 0.27 $\pm$ 0.02 & 103.1 $\pm$ 5.1 & 0.913 $\pm$ 0.080 & 4431 $\pm$ 388 & $\cdots$ & $\cdots$ & Fe~K~plume \\
270 & 128.04 & 103.5 & 0.36 $\pm$ 0.02 & 104.2 $\pm$ 3.8 & 0.950 $\pm$ 0.063 & 5730 $\pm$ 382 & +753 $\pm$ 5 & 55 & Fe~K~plume \\
272 & 149.54 & 107.4 & 0.42 $\pm$ 0.02 & 107.9 $\pm$ 3.4 & 0.954 $\pm$ 0.056 & 6724 $\pm$ 391 & -1413 $\pm$ 6 & 56 & Fe~K~plume \\
273 & 151.90 & 105.0 & 0.41 $\pm$ 0.02 & 105.8 $\pm$ 3.4 & 0.912 $\pm$ 0.054 & 6529 $\pm$ 389 & $\cdots$ & $\cdots$ & Fe~K~plume \\
274 & 110.61 & 110.1 & 0.28 $\pm$ 0.03 & 113.3 $\pm$ 5.1 & 0.874 $\pm$ 0.080 & 4557 $\pm$ 415 & $\cdots$ & $\cdots$ & Fe~K~plume \\
275 & 115.89 & 112.0 & 0.37 $\pm$ 0.02 & 115.7 $\pm$ 3.8 & 1.080 $\pm$ 0.071 & 5900 $\pm$ 386 & +14 $\pm$ 11 & 59, 60 & Fe~K~plume \\
277 & 123.94 & 110.5 & 0.31 $\pm$ 0.02 & 108.0 $\pm$ 4.5 & 0.863 $\pm$ 0.067 & 5042 $\pm$ 391 & -1651 $\pm$ 4 & 58 & Fe~K~plume \\
278 & 133.77 & 109.6 & 0.36 $\pm$ 0.02 & 109.0 $\pm$ 3.8 & 0.931 $\pm$ 0.063 & 5866 $\pm$ 394 & $\cdots$ & $\cdots$ & Fe~K~plume \\
279 & 163.41 & 112.2 & 0.46 $\pm$ 0.02 & 115.1 $\pm$ 3.0 & 0.954 $\pm$ 0.050 & 7348 $\pm$ 387 & $\cdots$ & $\cdots$ & Fe~K~plume \\
282 & 83.10 & 115.3 & 0.24 $\pm$ 0.03 & 120.1 $\pm$ 5.8 & 0.998 $\pm$ 0.104 & 3908 $\pm$ 407 & -674 $\pm$ 3 & 61 & Fe~K~plume \\
287 & 173.35 & 112.5 & 0.51 $\pm$ 0.02 & 114.0 $\pm$ 2.7 & 1.005 $\pm$ 0.048 & 8213 $\pm$ 388 & $\cdots$ & $\cdots$ & Fe~K~plume \\
291 & 110.21 & 124.9 & 0.29 $\pm$ 0.03 & 127.6 $\pm$ 5.3 & 0.891 $\pm$ 0.086 & 4629 $\pm$ 447 & $\cdots$ & $\cdots$ & Fe~K~plume \\
293 & 127.97 & 128.7 & 0.34 $\pm$ 0.02 & 128.8 $\pm$ 4.1 & 0.910 $\pm$ 0.065 & 5485 $\pm$ 394 & -44 $\pm$ 7 & 65 & Fe~K~plume \\
295 & 102.51 & 133.9 & 0.27 $\pm$ 0.02 & 135.4 $\pm$ 5.1 & 0.889 $\pm$ 0.080 & 4295 $\pm$ 384 & $\cdots$ & $\cdots$ & Fe~K~plume \\
299 & 187.27 & 142.1 & 0.54 $\pm$ 0.02 & 142.5 $\pm$ 2.5 & 0.993 $\pm$ 0.044 & 8764 $\pm$ 385 & +1957 $\pm$ 2 & 67 & Fe~K~plume \\
325 & 98.75 & 330.8 & 0.24 $\pm$ 0.02 & 334.8 $\pm$ 5.7 & 0.825 $\pm$ 0.082 & 3837 $\pm$ 382 & $\cdots$ & $\cdots$ & Northern Area \\
327 & 110.00 & 358.5 & 0.30 $\pm$ 0.02 & 0.4 $\pm$ 4.5 & 0.932 $\pm$ 0.074 & 4832 $\pm$ 381 & $\cdots$ & $\cdots$ & Northern Area \\
345 & 104.86 & 93.3 & 0.28 $\pm$ 0.02 & 89.6 $\pm$ 4.8 & 0.913 $\pm$ 0.077 & 4513 $\pm$ 381 & $\cdots$ & $\cdots$ & Fe~K~plume \\
\enddata
\tablecomments{(1) Knot ID; (2) projected angular distance from the explosion center; (3) position angle; (4) proper motion magnitude; (5) proper motion direction; (6) expansion index; (7) proper motion velocity assuming a distance of 3.4~kpc; (8) LOS velocity from \citet{koo23}; (9) corresponding knot numbers in \citet{koo23}; (10) classification based on position angle (see Section~\ref{ssec:decel}).}
\end{deluxetable}

\subsection{Deceleration of Outer FMKs \label{ssec:decel}}

\begin{figure}
\fig{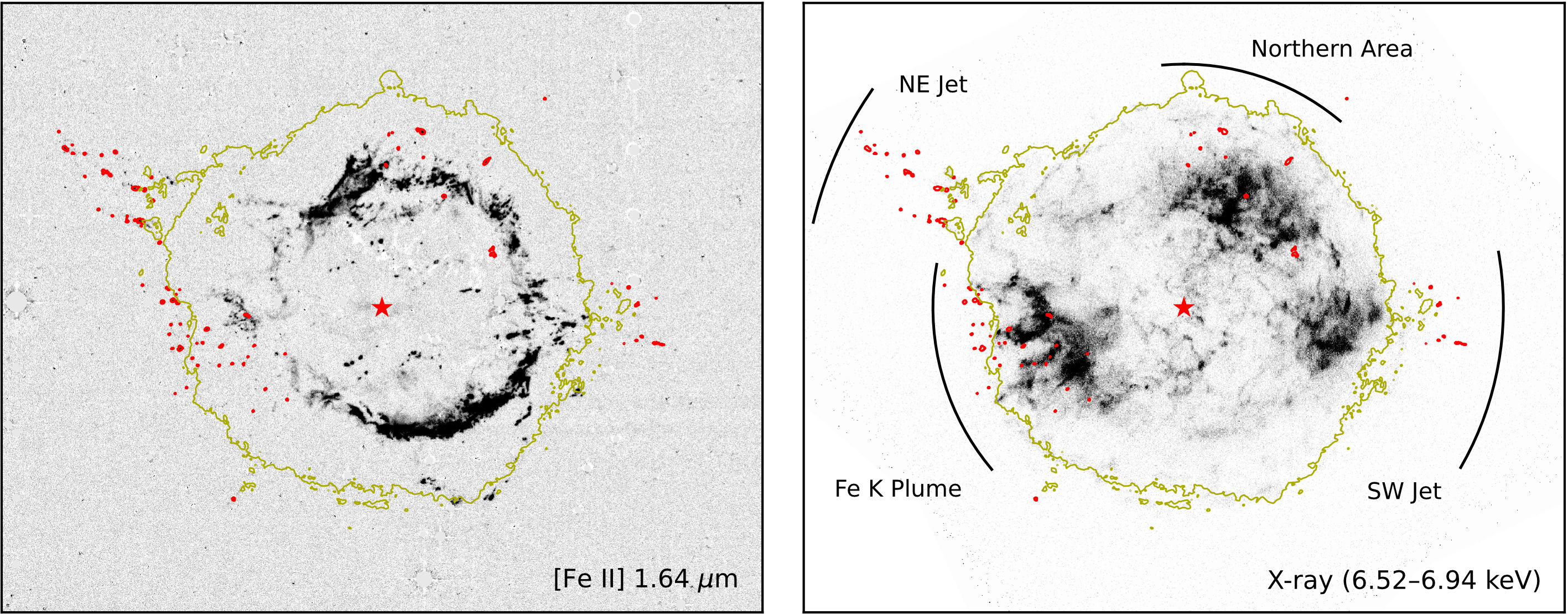}{0.9\linewidth}{}
\caption{\emph{Left:} Distribution of the identified FMKs in the 2020 \img. Red contours indicate the positions of the FMKs. North is up, and east is to the left. The dark yellow contour marks the approximate boundary of the SNR in radio in 2003 \citep{delaney04}. \emph{Right:} Same FMKs overlaid on the {\textrm{Chandra}} X‐ray image in the 6.52--6.94 keV band. The FMKs are divided into four groups for reference. \label{fig:feii_xray}}
\end{figure}

\begin{figure}
\fig{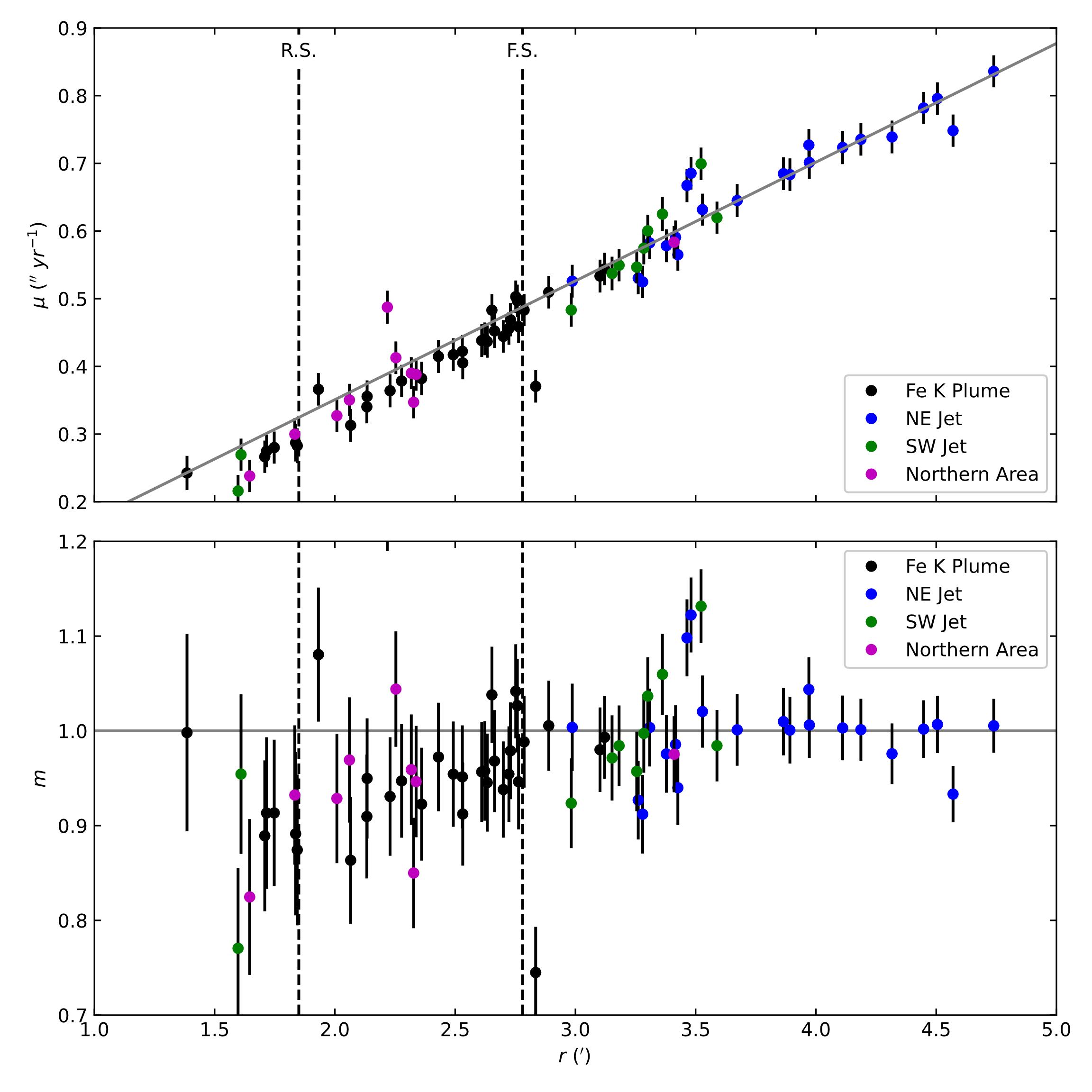}{0.8\linewidth}{}
\caption{Proper motion ($\mu$; upper panel) and expansion index ($m$; lower panel) of FMKs in Cas~A as functions of projected angular distance ($r$) from the explosion center. In both panels, the gray line indicates free expansion, and the vertical dashed lines mark the positions of the forward and reverse shocks. The symbol colors denote the FMK classes (see Section~\ref{ssec:decel}). \label{fig:mplot}}
\end{figure}

Proper motions of the FMKs located outside the main shell have been examined in several previous optical studies \citep{kamper76,vandenbergh83,thorstensen01,fesen06}. Among these, \citet{thorstensen01} conducted a detailed study of the proper motions of 38 FMKs using data obtained from both ground telescopes and \hst\ spanning the period from 1951 to 1999. By averaging the proper motions of 17 compact and morphologically stable outer knots under the assumption of no deceleration, they derived an explosion date of 1671.3$\pm$0.9. However, three of these knots (designated as Nos. 1, 4, and 118) yielded significantly earlier explosion dates, suggesting that they may have experienced measurable deceleration (see their Figure~5). \citet{fesen06} identified 1825 outlying ejecta knots at projected distance from the explosion center $r>100\arcsec$ using 2004 \hst\ images taken nine months apart and measured their proper motions. Based on the full sample of knots and assuming free expansion, they derived an average explosion year of 1662$\pm$27. When restricting their analysis to a selected subsample of 72 bright and compact knots, they obtained 1672$\pm$18, in agreement with the result of \citet{thorstensen01}. The earlier explosion date derived from the full sample corresponds to an average expansion index of $m=0.974\pm 0.079$ when adopting the explosion year of 1671.3, implying that many of the knots may have undergone modest but measurable deceleration. This finding supports the presence of differential deceleration among the outer ejecta, although the observed scatter could also reflect a combination of proper-motion uncertainties and intrinsic dispersion in knot velocities.

To facilitate discussion of deceleration among the outer FMKs, we divide them into four groups according to their position angles: NE~jet FMKs ($\theta=50\degr$--$78\degr$), Fe~K~plume FMKs ($78\degr$--$150\degr$), SW~jet FMKs ($230\degr$--$300\degr$), and northern area FMKs ($320\degr$--$10\degr$). Figure~\ref{fig:feii_xray} shows the locations of these FMK groups. The NE~jet FMKs consists of several narrow streams of compact knots extending well beyond the main shell, representing one of the most prominent outflows in Cas~A. The Fe~K~plume FMKs are those located around and outside the diffuse X-ray Fe~K~plume, an extended structure of X-ray-bright, Fe-rich ejecta gas that extends beyond the main ejecta shell and reaches nearly to the forward shock front \citep[; see Figure~\ref{fig:feii_xray}]{hwang12}. The Fe~K emission in Cas~A exhibits a highly asymmetric morphology, consisting of several large plumes that extend to or beyond the main shell, with the southeastern Fe~K~plume being the most prominent. The SW~jet FMKs, although fainter than those in the northeast, trace a broader, less collimated counter-jet structure that extends beyond the southwestern rim. The full distribution of this knot group spans a wider range of position angles, and the FMKs shown in Figure~\ref{fig:feii_xray} represent only a portion of the entire structure \citep[e.g., see Figure~8 of ][]{milisavljevic13}. Finally, the northern area FMKs comprise a sparse collection of outer ejecta knots located just outside the crown-like ring structure in the main ejecta shell. 

Figure~\ref{fig:mplot} shows our results, showing the proper motions and expansion index of FMKs as a function of their distance from the explosion center. The solid lines in the upper and lower panels indicate the relation expected for the free expansion, corresponding to $\mu = r/t$ or $m=1$. Most knots exhibit proper motions broadly consistent with free expansion. However, the knots located in the Fe~K~plume exhibit systematically low $m$ values. The mean $m$ values (after 3$\sigma$-clipping) are: 0.964$\pm$0.010 for Fe~K~plume, 0.998$\pm$0.008 for NE~jet, 0.974$\pm$0.019 for Northern Area, and 1.008$\pm$0.014 for SW~jet. These results indicate that knots in the Fe~K~plume have undergone measurable deceleration, while those in the NE~jet remain closest to pure ballistic expansion. The knots in the Northern Area and SW~jet show marginal evidence of deceleration, though with larger uncertainties due to their smaller sample sizes. 

As demonstrated in Figure~\ref{fig:mplot}, knots in the Fe~K region and the northern area are more decelerated than those in the jet region. This finding is consistent with the observed deceleration of ``shell'' knots, most of which are located in the northern area, as discussed in \citet{thorstensen01}. However, their ``shell'' knots and ``outer'' knots demonstrate a discrepancy of nine years in the explosion date, which is less than the discrepancy of 14 years between our Fe~K~plume and NE~jet knots. This discrepancy may arise from differences in the specific knots that were tracked. When matching our catalog to theirs, we identified 12 common knots. Among these, five inner knots with proper motions less than $0\farcs5$ yr$^{-1}$ decelerated 6.3\% more on average than those in \citet{thorstensen01}, whereas seven outer knots show a slight increase of 1.4\%. This contrast highlights the enhanced deceleration of the inner knots. Nevertheless, given that the sample sizes are limited and their discrepancy exceeds the average differences, the statistical significance of this difference remains uncertain.

The Fe~K~plume region of Cas~A is distinguished by Fe-rich ejecta that are bright in X-ray Fe~K emission \citep{willingale02, hwang03, hwang04, hwang12}. It appears as a prominent, finger-like protrusion extending well beyond the main shell, implying that Fe-rich material synthesized in the innermost, high-entropy regions during the explosion expanded faster than the bulk ejecta and penetrated through the overlying Si- and O-rich layers \citep[see Figure~\ref{fig:feii_xray}; ][]{delaney10, milisavljevic13}. Numerical simulations indicate that large-scale Fe-rich plumes originate from stochastic processes such as neutrino-driven convection and the standing accretion shock instability during the first seconds after core collapse, while Rayleigh-Taylor instabilities shape their subsequent evolution \citep{wongwathanarat17}. These plumes begin interacting with the reverse shock within a few decades after the explosion, leading to measurable deceleration \citep{orlando21}. Observationally, the X-ray Fe-rich plasma exhibits transverse velocities of $\sim$4300--6700\kms, corresponding to an expansion index of $m = 0.67$--$0.86$ \citep{tsuchioka22}.

The chemical and kinetic properties of dense FMKs in the Fe~K~plume area were examined in detail by \citet{koo23}. They found that the knots along the outer boundary of the diffuse X-ray-emitting Fe~K~plume are Fe-rich, exhibiting only [Fe II] emission lines with no detectable [S II] lines. These Fe-rich knots show transverse velocities of 6600--7600\kms, higher than those of the diffuse Fe-rich plasma (4300--6700\kms; \citealt{tsuchioka22}), though their radial velocities are comparable ($-2600$ to $-500$\kms). From these spatial and kinematic correlations, \citet{koo23} concluded that the compact Fe-rich knots are physically associated with the diffuse Fe ejecta.

Our measurements show that the Fe-plume knots have undergone only modest deceleration during their interaction with the reverse shock and the surrounding medium. Their present expansion index of $m =$0.96 corresponds to just a $\sim$4\% reduction in velocity relative to free expansion, implying that they have largely retained their initial ballistic motion. This behavior stands in contrast to the surrounding diffuse X-ray-emitting Fe-rich plasma, which shows significantly lower expansion indices of $m =$0.67--0.86, indicative of stronger deceleration. One possible interpretation is that the dense FMKs, owing to their compact sizes and high column densities, are able to preserve their ballistic motion more effectively than the diffuse Fe-rich ejecta. In this context, it is worth noting that numerical simulations of Cas~A \citep[e.g.,][]{orlando16,orlando21,orlando25b} predict that large-scale Fe-rich ejecta plumes produced in neutrino-driven explosions encounter the reverse shock at early times ($\sim$tens of years after the explosion). In these models, hydrodynamic instabilities at the contact discontinuity between shocked ejecta and shocked circumstellar material can form dense ejecta clumps. Such clumps are expected to cool rapidly and may become observationally faint, but could potentially be re-illuminated if they later interact with denser circumstellar material outside the forward shock. The apparent concentration of FMKs near the outer boundary of the diffuse Fe~K~plume is qualitatively consistent with such an evolutionary picture. 

\section{Expansion of Main Ejecta Shell \label{sec:shell}}

\subsection{Proper Motion Measurement \label{ssec:proper}}

We measured the proper motions of compact features in the main ejecta shell using cross-correlation, following the approach described in Section~\ref{sssec:fmkpm}. First, as outlined in Section~\ref{ssec:identify}, we generated contour masks from the 2013 \img\ using a threshold of $4\times 10^{-18}$ \unit\ pixel$^{-1}$. From these, we excluded features outside the shell, such as identified knots, stellar residuals, and observational artifacts, to construct the shell mask.

It has been determined that certain regions of the mask lack compact features and local minima, which are essential for determining precise positional movements. To exclude these, the contour threshold was increased by 1\% until the mask area reached the maximum (25 arcsec$^{2}$) or the mask was divided into pieces. The maximum area threshold was set to split shell features from nearby QSFs while excluding false detections from the local correlation peak of small contours. The minimum area value was increased from five pixels (0.2 arcsec$^{2}$; see Section~\ref{ssec:identify}) to 25 pixels (1 arcsec$^{2}$), ensuring reliable tracking via cross-correlation by containing full compact features in the mask. The total number of mask regions following this procedure is 515.

We determined the shifts of each mask using a discrete Fourier transform upsampling technique, which pads with zeros to account for unmeasured high-frequency components in the Fourier domain \citep{guizar08}. The upsampling factor was set to five, resulting in a pixel size of 0\farcs04. Cross-correlation maps were generated from the upsampled 2013 and 2020 images, with a search range extending up to 4\arcsec. Since the knots are only mildly decelerated, we searched for a local peak within a circular area centered on the position corresponding to 90\% of the free expansion shift and with a radius equal to 20\% of the magnitude of the free expansion shift. This method works for most FMKs, but in some cases, the local peaks are not clearly defined. Additionally, static artifacts or QSFs can complicate the results. To address this, we also search for another local peak within a circular area of the same radius, but centered on the position with no shift. The final result is taken from the first peak when its correlation coefficient is higher than that of the static peak. This approach excludes masks with no peak at the estimated position or those with a higher correlation at the shiftless position. The former masks were primarily darkened features not visible in 2020, while the latter were mainly QSFs or static artifacts. In total, the analysis yielded 215 measurements.

\begin{figure}
\fig{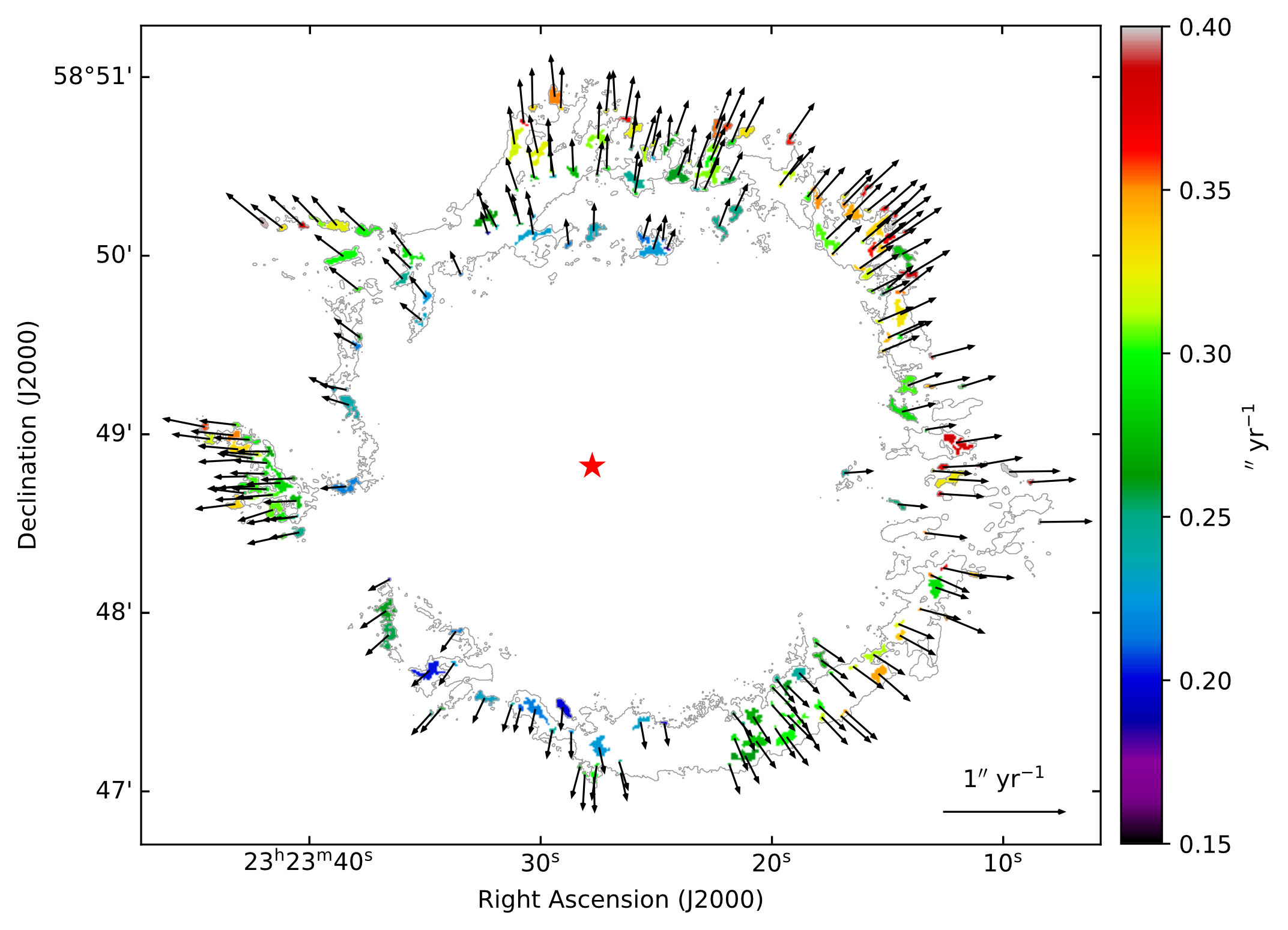}{0.9\textwidth}{}
\caption{Proper motion map of the ejecta shell. Each knot is color-coded by its proper motion magnitude, and black arrows indicate the corresponding motion vectors. A scale bar of 1\arcsec\ yr$^{-1}$ is shown at the bottom right. The red star marks the explosion center. A gray contour outlines the main ejecta shell. \label{fig:pmmap}}
\end{figure}

Figure~\ref{fig:pmmap} illustrates the proper motions of individual knot features in the main ejecta shell. The magnitude of motion is shown by a color gradient, ranging from black through purple to red, with warmer colors indicating faster motions. Black arrows mark the direction of motion. The knot proper motion ranges from 0\farcs18 yr$^{-1}$ to 0\farcs43 yr$^{-1}$ (2900--6900\kms), with a median value of 0\farcs3 yr$^{-1}$ (4800\kms). Most knots appear green, corresponding to the median velocity, whereas those along the northern outer rim exhibit larger proper motions and appear red. In contrast, knots in the southeastern region show smaller proper motions and appear blue. The knots in the northern inner region, which exhibit large LOS velocities \citep[e.g., see Figure~5 of ][]{milisavljevic13}, also have smaller proper motions and appear blue.

\subsection{Expanding Spherical Shell Model \label{ssec:linmodel}}

The main ejecta shell of the Cas~A appears as an approximately circular ring when projected onto the sky. We modeled this structure as a circular ring with a Gaussian radial brightness profile and determined its geometric parameters by fitting the model to the observed brightness distribution in the 2013 image, which has higher sensitivity. The fitting was performed using a least-squares optimization implemented with the \texttt{curve\_fit} routine in the \texttt{SciPy} Python package. The errorbars of each parameter were obtained from the reduced $\chi^2$.

The initial parameters for the fit were chosen as follows: the ring center was set at the center of the optical nebulosity given by \citet{minkowski59}, (23$^{\rm h}$ 23$^{\rm m}$ $25.2^{\rm s}$, +58$^\circ$ 48\arcmin\ 55\arcsec)$_{\rm J2000}$, the amplitude was set to the peak intensity of the image, the ring radius was set to 1\farcm7 which is consistent with the observed shell structure, and the Gaussian width was set to 20\arcsec. To minimize contamination from noise, outer knots, and unrelated background features, we employed an iterative fitting procedure that retained only the pixels lying within a $3\sigma$ boundary of the ring in each iteration, thereby converging on a representation that closely traces the true shell morphology. We applied the QSF mask defined in Section \ref{ssec:identify} to exclude QSFs from the fit. We checked the fitting with varied initial parameters and confirmed convergence to consistent parameter values.

\begin{figure}
\fig{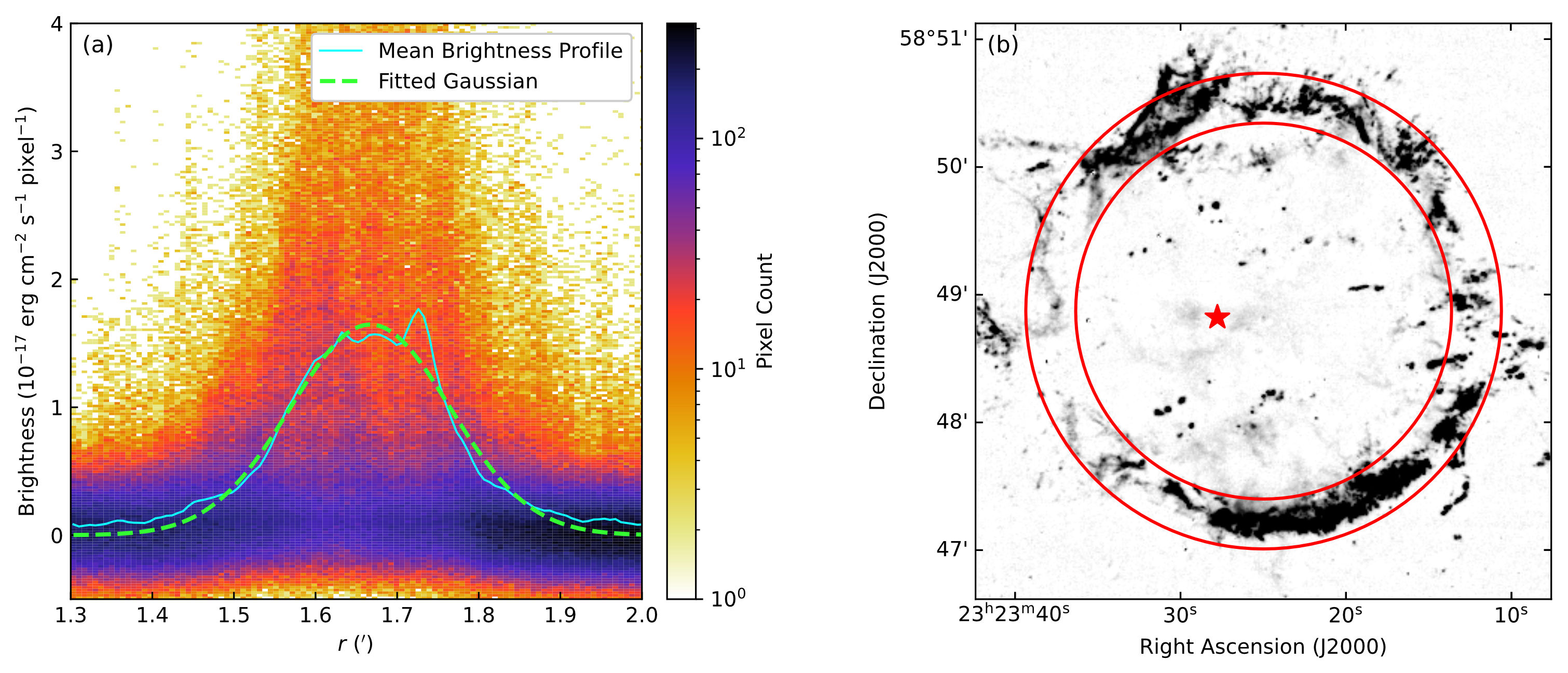}{0.85\textwidth}{}
\caption{Result of the Gaussian ring model fit.
\emph{Left:} Comparison of the fitted model and observed data. The background image shows the 2D histogram of pixel values in the radial distance-brightness plane. The white line shows the image's mean brightness profile, and the green dashed line shows the fit result. The projected radial distance was calculated from the fitted center. QSFs are masked in this figure. 
\emph{Right:} Gaussian ring fit result on the image. The background is 2013 image with grayscale from $-2\times10^{-18}$ \unit\ pixel$^{-1}$ to $2\times10^{-17}$ \unit\ pixel$^{-1}$. Two red circles show the 2-$\sigma$ range of the Gaussian ring fit. The red star marks the explosion center. \label{fig:ring}}
\end{figure}

Figure~\ref{fig:ring} shows the fit result. The model reproduces the radial profile of the mean brightness reasonably well and, in the image plane, traces the prominent northern and southern arcs of the shell, indicating that these bright features largely define the overall ring structure. The derived geometric parameters are summarized in Table~\ref{tab:shellparameter}. The best-fit coordinate of the geometric center of the ejecta ring is found to be (23$^{\rm h}$ 23$^{\rm m}$ $24.99^{\rm s}\pm 0.06^{\rm s}$, +58$^\circ$ 48\arcmin\ $52\farcs2\pm0\farcs5$)$_{\rm J2000}$, offset from the explosion center of \citet{thorstensen01} by $(-21\farcs7,2\farcs8)$. We emphasize that the geometric center does not correspond to a location where the proper motion vectors vanish; rather, it is a purely morphological construct that represents the centroid of the projected brightness distribution in the plane of the sky.

\begin{deluxetable}{lc}
\tablecaption{Global Parameters of the Main Ejecta Shell \label{tab:shellparameter}}
\tablewidth{0pt}
\tablehead{\colhead{Parameter} & \colhead{Value}}
\startdata
Geometric center & $(23^{\rm h}23^{\rm m}24\fs99 \pm 0\fs06,\ +58\arcdeg48\arcmin52\farcs2 \pm 0\farcs5)_{\rm J2000}$ \\
Offset from explosion center$^{\ast}$ & ($-$21.7$\pm$0.5\arcsec , 2.8$\pm$0.5\arcsec ) ($-$0.36, 0.05) pc \\
Radius & 100.0$\pm$0.3\arcsec\ (1.65 pc) \\
Radial thickness & 11.7$\pm$0.3\arcsec\ (0.19 pc) \\
Systemic proper motion & ($-$0.0648 $\pm$ 0.0042, 0.0130 $\pm$ 0.0032) \arcsec\ yr$^{-1}$ \\
 & ($-$1044 $\pm$ 67, 210 $\pm$ 52)\kms \\
Radial proper motion & 0.2746 $\pm$ 0.0026 \arcsec\ yr$^{-1}$ \\
 & (4425 $\pm$ 42\kms) \\
Expansion rate & 0.2746 $\pm$ 0.0026 \% yr$^{-1}$ \\
Expansion timescale & 364 $\pm$ 3 yr \\ 
Expansion index$^{\ast}$ & 0.940 $\pm$ 0.009 \\
\enddata
\tablecomments{Secondary values in units of pc or \kms\ assume a distance of 3.4 kpc. \\
$^{\ast}$ The explosion center and date are adopted from \citet{thorstensen01}: \cen $_{\rm J2000}$ and 1671.3$\pm$0.9. \\
}
\end{deluxetable}

\begin{figure}
\fig{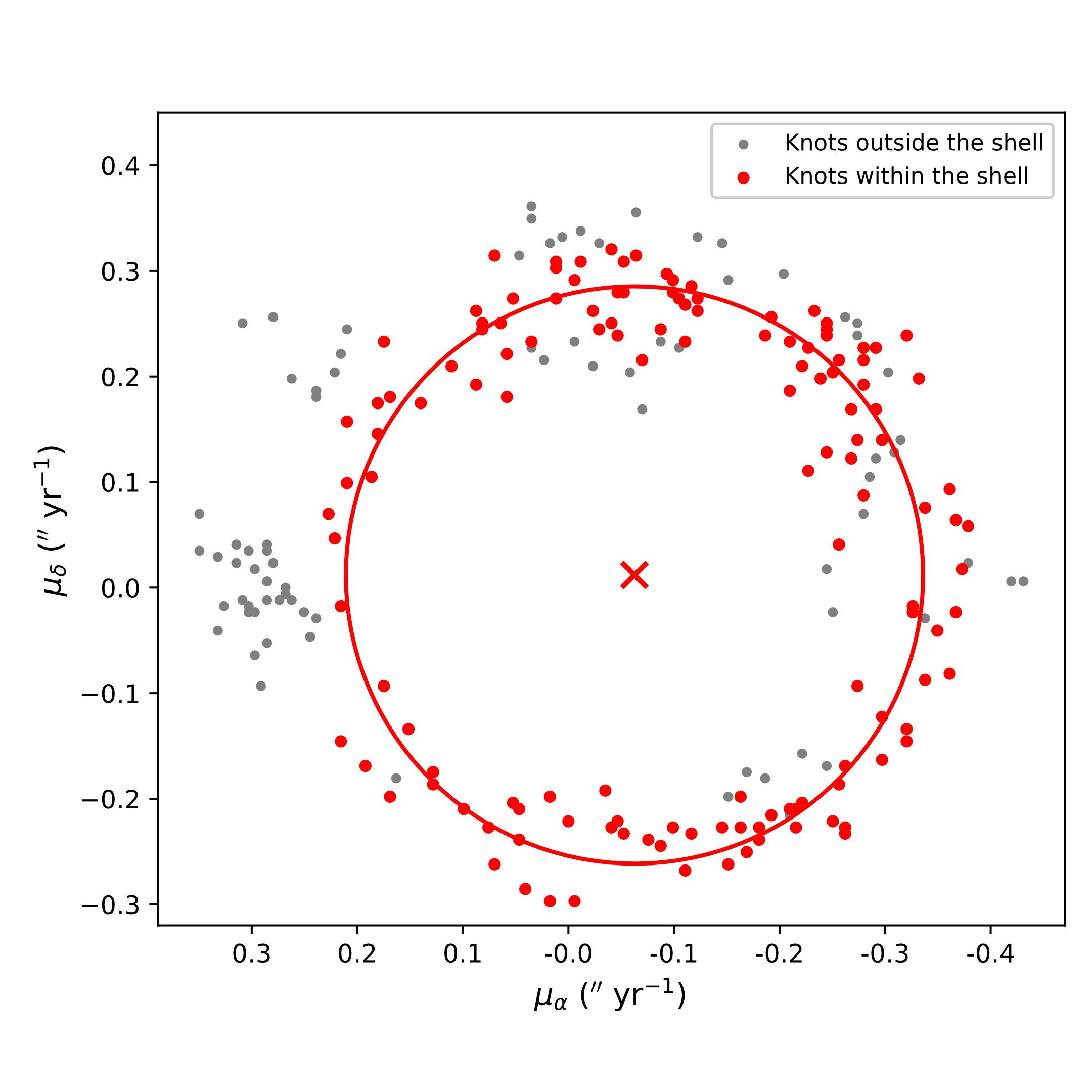}{0.45\textwidth}{}
\caption{Measured proper motions in R.A. and Dec. for knots in the main ejecta shell. Knots that lie within the 2-$\sigma$ range of the Gaussian ring model are highlighted in red. The red circle and its center mark the least-squares best-fit circle to all data points.}
\label{fig:muxmuy}
\end{figure}

Figure~\ref{fig:muxmuy} shows the R.A. and Dec. components ($\mu_{\alpha}$, $\mu_{\delta}$) of the compact features in the main ejecta shell. The distribution is well described by a circular ring in proper-motion space, and we fit the data in the $2-\sigma$ Gaussian ring region with a model of a uniformly expanding ring with a bulk systemic motion: 
\begin{equation}
(\mu_\alpha-\mu_{\alpha,0})^2 + (\mu_\delta-\mu_{\delta,0})^2 = \mu_{\rm exp}^2
\end{equation}
where $(\mu_{\alpha,0}, \mu_{\delta,0})$ is the center of the locus in the proper motion space (the systemic motion) and $\mu_{\rm exp}$ is the characteristic radial proper motion of the shell. The best-fit parameters are $\mu_{\alpha,0}=-0\farcs0648 \pm 0\farcs0042$ yr$^{-1}$ and $\mu_{\delta,0}=0\farcs0130 \pm 0\farcs0032$ yr$^{-1}$, which correspond to $-1044$\kms\ and $+210$\kms\ at $d=3.4$~kpc, respectively. The radial proper motion is $\mu_{\rm exp}= 0\farcs2746 \pm 0\farcs0026\ \mathrm{yr^{-1}}$ ($4425 \pm 42$\kms). The quoted uncertainties are 1$\sigma$ errors estimated from fitting.

\begin{figure}
\fig{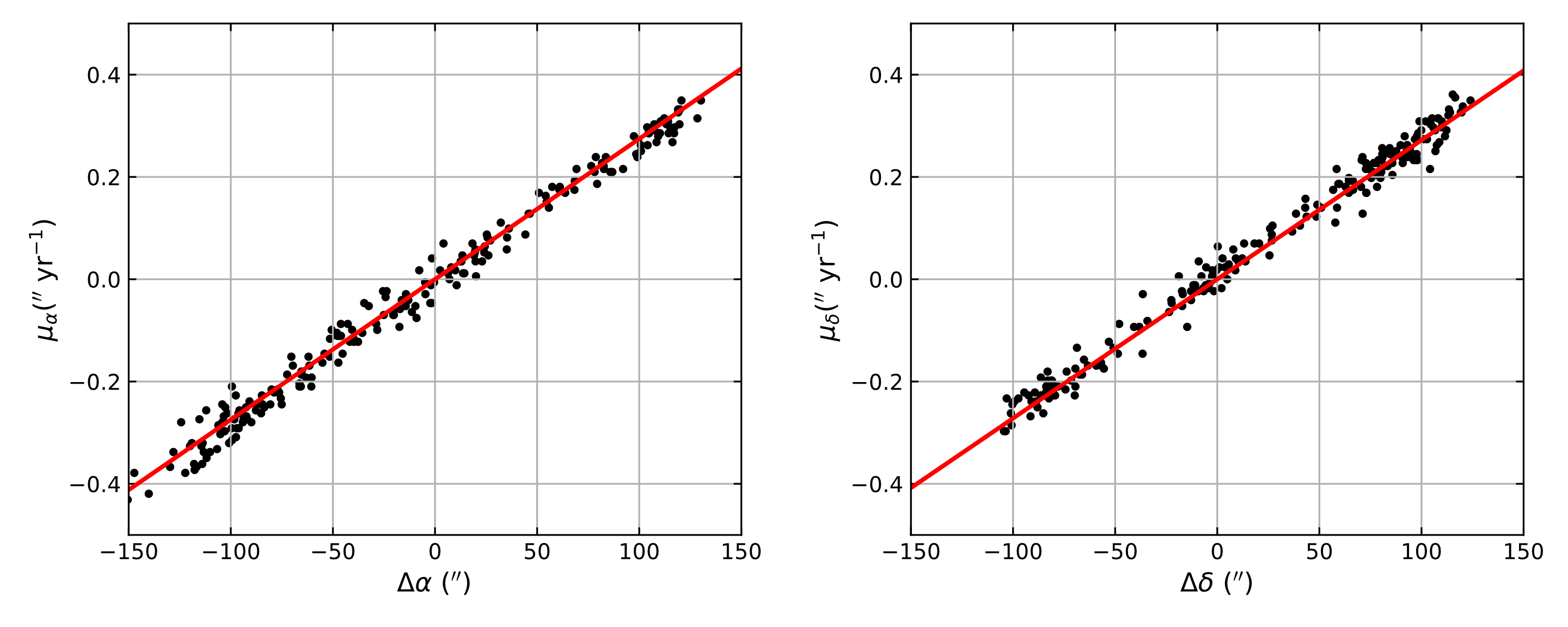}{0.9\textwidth}{}
\caption{Comparison between the proper motions of main shell ejecta knots and their projected displacements from the explosion center along right ascension (left) and declination (right). The red lines denote the best-fit linear relations.}
\label{fig:linfit}
\end{figure}

We also examined $\mu$ as a function of the projected distance from the explosion center along the RA and Dec directions (Figure~\ref{fig:linfit}). Both axes exhibit tight linear trends. Using a 3$\sigma$-clipped linear fit with the center fixed at the explosion center, we obtained slopes of $(2.745 \pm 0.021)$ and $(2.719 \pm 0.022)\times10^{-3}$ \%~yr$^{-1}$ along the RA and Dec, respectively. These slopes correspond to an expansion time scale of $\sim 365$~yr, consistent with the value derived from the mean radius and radial proper motion in Table~\ref{tab:shellparameter}. 

The expansion parameters derived here are compared to those of earlier works in Table~\ref{tab:mcompare}. Only a few works have investigated the expansion of the main ejecta shell in the plane of the sky. \citet{kamper76} measured the proper motions of about one hundred optical knots using Palomar plates obtained between 1951 and 1975, deriving an expansion rate of 0.324\% yr$^{-1}$. \citet{thorstensen01} analyzed 21 main shell knots observed from 1951 to 1999 with both ground-based telescopes and the \hst, obtaining an expansion rate of 0.296\% yr$^{-1}$. Without accounting for deceleration, these rates correspond to expansion ages of $\simeq309$~yr and $\simeq338$~yr, respectively, whereas our best-fit rate of $\simeq 0.275\%$~yr$^{-1}$ implies an age of $\simeq364$~yr. This difference may partly reflect differences in time baselines and the transient nature of optical knots, which can become visible when shocked, fade over time, and be followed by newly shocked knots.

These early optical measurements were largely limited to knots in the northern and eastern regions, because the southern portion of the shell was faint or absent at that time. More recently, \citet{fesen25} analyzed multiepoch \hst\ images of bright optical features around the main shell and measured proper motions of $0\farcs216$--$0\farcs368$ yr$^{-1}$, with a mean of $0\farcs294\pm0\farcs011$ yr$^{-1}$ (4736 $\pm$ 183\kms\ at 3.4 kpc). The corresponding expansion rate (0.294\% yr$^{-1}$) is almost identical to that of \citet{thorstensen01}. They further found that knots along the western--northwestern rim expand systematically faster than those on the eastern--southeastern side, i.e., 0\farcs306--0\farcs368 yr$^{-1}$ versus 0\farcs216--0\farcs271 yr$^{-1}$. Given that our model assumes a spherically expanding shell with a bulk (systemic) motion, the slightly smaller expansion rate derived from our ring model, $\mu_{\rm exp}=0\farcs2746\pm0\farcs0026$ yr$^{-1}$ (4425 $\pm$ 42\kms), is reasonable. The inferred systemic motion of $\sim0\farcs066$ yr$^{-1}$ toward the northwest is also consistent with the west-northwest versus east-southeast asymmetry in the knot motions. This large-scale asymmetry in the shell's expansion is discussed further in the next section.

The derived radial proper motion of the main shell ejecta, $\mu=0\farcs2746\ \mathrm{yr^{-1}}$, is substantially below the free expansion rate expected for undecelerated ejecta, $R_s/t \approx 0\farcs292\ \mathrm{yr^{-1}}$ (that is, $0.292\%~\mathrm{yr^{-1}}$ for an explosion in 1671.3). This indicates measurable deceleration of the main shell. Assuming a power-law expansion ($r \propto t^m$), the implied expansion index is $m=0.94$. In other words, the main shell ejecta knots have slowed by $\sim 6$\% relative to free expansion. Physically, the knots likely expanded nearly freely until encountering the reverse shock, where they experienced compression and deceleration due to interaction with the shock and the material immediately upstream and downstream of it \citep[see also ][]{fesen06,fesen11}.

By contrast, the derived expansion rate of the main shell is significantly higher than that of the diffuse, X-ray-emitting ejecta (see Table~\ref{tab:mcompare}). If we adopt the shell radius of $R_s=100\arcsec$, the radial proper motion of the shell translates to a fractional expansion rate of $0.2746\% \pm 0.0026\%~\mathrm{yr^{-1}}$, whereas the expansion rate of the X-ray emitting ejecta is $\sim 0.20$\% $\mathrm{yr^{-1}}$ \citep{koralesky98,vink98}. This difference is expected: the ejecta comprise dense clumps embedded in a lower-density component that dominates the X-ray emissivity. Both components may travel freely until they reach the reverse shock; thereafter, the diffuse component suffers greater deceleration and is heated to X-ray temperatures, while the dense clumps experience a much slower internal shock and primarily radiate at optical/IR wavelengths. In the sky frame, reverse-shock speeds along the main shell inferred from X-ray analyses are $\sim$2600--4800\kms\ \citep[e.g.,][]{vink98, vink22, wu24, suzuki25}; see also Table~3 of \citet{fesen25}, whereas \citet{fesen25} typically find lower values of $\sim$1000--2000\kms. Thus, diffuse ejecta expanding at $\sim$5000\kms\ encounter an effective shock of several thousand\kms\ or more (in the ejecta frame), leading to substantial deceleration and strong X-ray emission. For dense clumps, the transmitted shock speed scales roughly as $v_{c}\sim v_X/\sqrt \chi$ for a clump-interclump density contrast $\chi$, yielding $v_c\lesssim$ 500\kms\ for plausible $\chi$ --- consistent with bright optical line emission and only modest deceleration of the clumps. In the western area of the SNR, however, optical and X-ray observations yield inconsistent estimates of the reverse-shock velocity (see Section~\ref{ssec:shelldecel}).

The metal-rich, dense ejecta knots heated by the reverse shock radiate for less than a few decades owing to their rapid radiative cooling \citep[e.g.,][]{raymond18,orlando25b}. Consequently, such knots are expected to lie just outside the reverse shock, farther from the explosion center, delineating a shell of recently shocked dense ejecta material. The mean radius of the main ejecta shell derived in this work, however, is $\sim100\arcsec$, smaller than the average reverse-shock radius of $\sim115\arcsec $ inferred from X-ray and radio measurements \citep[Table~\ref{tab:geometric}; ][]{helder08,arias18}. This apparent discrepancy likely arises from projection effects, as many knots are seen in projection along the LOS within the projected reverse-shock circle rather than located precisely on its three-dimensional surface \citep[e.g.,][]{vink22}. However, if the reverse-shock front is locally distorted by hydrodynamic instabilities --- such as the Richtmyer-Meshkov instability --- the ejecta knots could be located closer to the explosion center than the surrounding reverse-shock surface.

\begin{deluxetable*}{lccccc}
\tablecaption{Estimates of Average Expansion Parameters of the Main Ejecta Shell \label{tab:mcompare}}
\tablehead{
\colhead{Reference} & \colhead{Measured Feature} & 
\colhead{Expansion Velocity} & \colhead{Expansion Rate} & \colhead{Expansion Index $m$} & \colhead{Expansion Time Scale} \\
\colhead{} & \colhead{} & \colhead{(\kms)} & \colhead{(\% yr$^{-1}$)} & \colhead{} & \colhead{(yr)}}
\startdata
\underline{In the sky plane} & & & & & \\
\citet{kamper76} & Optical knots & $\cdots$ & $0.324 \pm 0.003$ & $\cdots$ & $308.6 \pm 2.9$ \\
\citet{thorstensen01} & Optical knots & $5628 \pm 265$ & $0.2963 \pm 0.0015$ & $0.973 \pm 0.005$ & $337.5 \pm 1.7$ \\
\citet{koralesky98}, \citet{vink98} & X-ray ejecta & $\sim 3500$ & $\sim 0.20$ & $\sim 0.73$ & $\sim 500$ \\
\citet{vink22} & X-ray ejecta & $2459 \pm 1540$ & $0.134 \pm 0.084$ & $0.446 \pm 0.279$ & $746 \pm 468$ \\
\citet{fesen25} & Optical knots & $4736 \pm 183$ & $0.294 \pm 0.004$ & $0.966 \pm 0.014$ & $340 \pm 5$ \\
\citet{suzuki25} & X-ray ejecta & $3966 \pm 186$ & $0.227 \pm 0.011$ & $0.80 \pm 0.04$ & $441 \pm 21$ \\
This Work & NIR \fe\ knots & $4425 \pm 42$ & $0.2746 \pm 0.0026$ & $0.940 \pm 0.009$ & $364 \pm 3$ \\
\\
\underline{Along the LOS$^*$} & & & & & \\
\citet{reed95} & Optical knots & $5290 \pm 90$ & $0.3141 \pm 0.0057$ & $0.991 \pm 0.018$ & $318.4 \pm 5.8$ \\
\citet{delaney10} & IR [\ion{Ar}{2}] ejecta & $4936 \pm 224$ & $0.282 \pm 0.026$ & $0.941 \pm 0.086$ & $354 \pm 32$ \\
\citet{milisavljevic13} & Optical knots & $4820 \pm 224$ & $0.2820 \pm 0.0384$ & $0.951 \pm 0.130$ & $354 \pm 48$ \\
\enddata
\tablecomments{Expansion indices $m$ are calculated using an explosion year of 1671.3 and observation dates given in each reference. The distance of 3.4kpc is used to convert parameters. For \citet{thorstensen01}, only shell knots were averaged with the last epoch 1999.79 as an observation date. For \citet{fesen25}, inverse-variance weighted mean values of Table~5 were used. For \citet{suzuki25}, we used the average of Table~4 values without region SE5, which is far from the main shell.}
\tablenotetext{*}{Doppler measurements along the LOS. The shell radius, or scaling factor, was used to convert the parameters.}
\end{deluxetable*}

\subsection{Gross Expansion Asymmetry}

\begin{deluxetable*}{lcccccc}
\tablecaption{Comparison of Main Ejecta Shell Geometric Parameters \label{tab:geometric}}
\tablehead{
\colhead{Reference} & \colhead{Measured Feature} & \colhead{Center} & 
\colhead{$\Delta r^\dagger$} & \colhead{$\theta^\dagger$} &
\colhead{Radius} & \colhead{Thickness} \\
\colhead{} & \colhead{} & \colhead{($\alpha$, $\delta$)$_{\rm J2000}$} & 
\colhead{(\arcsec)} & \colhead{(\degr)} & 
\colhead{(\arcsec)} & \colhead{(\arcsec)}
}
\startdata
\citet{minkowski59} & Optical, ejecta shell
 & (23$^{\mathrm{h}}$23$^{\mathrm{m}}$25\fs2,\ +58\arcdeg48\arcmin55\arcsec)
 & 20.8 & 285.6 & $\cdots$ & $\cdots$\\
\citet{reed95} & Optical, ejecta shell
 & (23$^{\mathrm{h}}$23$^{\mathrm{m}}$26\fs60,\ +58\arcdeg49\arcmin00\farcs5)
 & 14.4 & 320.7 & $104.5 \pm 0.7$ & 10\\
\citet{gotthelf01} & X-ray, reverse shock
 & (23$^{\mathrm{h}}$23$^{\mathrm{m}}$25\fs44,\ +58\arcdeg48\arcmin52\farcs3)
 & 18.3 & 279.1 & $95 \pm 10$ & $\cdots$ \\
\citet{helder08} & X-ray, ejecta shell
& $\cdots$
 & $\sim$ 15 & $\sim$ 270 & $\sim 115$ & $\sim 30$ \\
\citet{arias18} & Radio, reverse shock
 & (23$^{\mathrm{h}}$23$^{\mathrm{m}}$26\fs00,\ +58\arcdeg48\arcmin54\arcsec)
 & 14.5 & 288.6 & $114 \pm 6$ & $\cdots$ \\
This work & NIR \fe, ejecta shell
 & (23$^{\mathrm{h}}$23$^{\mathrm{m}}$24\fs99,\ +58\arcdeg48\arcmin52\farcs24)
 & 21.8 & 277.4 & $100.0 \pm 0.3$ & $11.7 \pm 0.3$ \\
\enddata
\tablenotetext{\dagger}{Offset from the explosion center of \citet{thorstensen01}: \cen $_{\rm J2000}$.} 
\end{deluxetable*}

We have found that the geometric center of the main ejecta shell is offset by about 22\arcsec\ (or 0.4 pc) toward the northwest (P.A.$\approx 280^\circ$) relative to the explosion center. This displacement is not new: it was first noted in early optical studies of ejecta kinematics \citep{reed95,thorstensen01} and has since been confirmed by multiwavelength work (Table~\ref{tab:geometric}). In X-rays, analyses of the reverse-shocked plasma have shown that the center of the reverse shock is similarly displaced to the northwest \citep{gotthelf01, helder08, vink22}, consistent with the bright main shell delineating ejecta that have passed through the reverse shock. The reverse shock location has also been independently inferred from low-frequency radio free-free absorption against the central synchrotron emission, which traces the distribution of unshocked cold ejecta interior to the reverse shock; these radio data likewise indicate a northwest displacement \citep{arias18}. 

The offset of the geometric center indicates an asymmetric expansion of the main shell. Our proper-motion fit yields a bulk motion of  $(\mu_{\alpha,0}, \mu_{\delta,0}) = (-0\farcs065, +0\farcs013)$~yr$^{-1}$, corresponding to ($-1044$, +210)\kms. The resulting net systemic velocity is $0\farcs066$~yr$^{-1}$ ($\approx1065$\kms) toward P.A.~$\approx280^\circ$. If this drift had persisted uniformly over 350~yr, the implied displacement would be $\approx0.38$~pc ($\approx$23\arcsec\ at 3.4~kpc), consistent with the geometric-center offset listed in Table~\ref{tab:shellparameter}. We note that the proper motions presented in this work were measured by tracking individual, shocked, dense [Fe II]-emitting features that retain similar morphologies over the observational baseline. The features that show substantial morphological evolution, likely associated with the emergence of newly shocked ejecta material and/or the fading of previously shocked features as the reverse shock propagates, were not included, since their proper motions cannot be measured reliably. We therefore interpret the derived expansion rates as tracing the bulk motion of the expanding ejecta, rather than a pattern speed.

For comparison, optical and infrared spectroscopic Doppler measurements indicate that the LOS velocities of the main shell ejecta span from $-4000$\kms\ to $+6000$\kms\ \citep{lawrence95,reed95,alarie14}. Model fits to these data suggest that the shell is redshifted with a mean receding velocity of $\sim$800\kms\ \citep{delaney10, milisavljevic13}. Correcting for Cas~A's systemic velocity of about $-50$\kms\ and allowing for a modest progenitor proper motion \citep{koo25}, the inferred global offset along the LOS remains essentially unchanged. Taken together, these results show that the main ejecta shell does not expand isotropically about the explosion center, but instead exhibits a coherent bulk displacement: toward the northwest in the sky plane and away from the observer along the LOS, with a net asymmetry of order $\sim 1300$\kms. 

It is also noteworthy that the expansion velocity along the LOS differs from that in the plane of the sky. Spectroscopic studies report mean LOS expansion velocities of 4800--4900\kms, which are noticeably higher than the expansion velocity measured in the sky plane in this study (see Table~\ref{tab:mcompare}). Since these spectroscopic analyses also assumed a spherically expanding shell with a systemic LOS motion, the higher LOS expansion velocity suggests that the main ejecta shell is expanding slightly faster along the LOS, indicating a modest elongation of the SNR toward the observer.

Several explanations have been advanced to account for the asymmetries in Cas~A. One possibility is that the SNR is expanding into a CSM or interstellar medium with a smooth density gradient. In such a configuration, the reverse shock is less developed on the lower-density side, causing the shocked ejecta to expand more rapidly in that direction. \citet{reed95} suggested that Cas~A is embedded in an environment where the density is lower on the redshifted far side, an interpretation supported by the lack of QSFs with redshifted velocities. An alternative explanation is that the asymmetry is intrinsic to the explosion itself: three-dimensional reconstructions of Cas~A show large-scale plumes and overturned layers that are difficult to reproduce without invoking directional asymmetries in the explosion mechanism \citep{delaney10, isensee10}. On the other hand, \citet{milisavljevic13} pointed out that the rear-facing ejecta appear to be truncated abruptly along an inclined plane, whereas the front-facing ejecta do not --- an observation more consistent with external environmental effects than with purely intrinsic explosion asymmetries.

Recent three-dimensional simulations of neutrino-driven SN explosions show that large-scale asymmetries naturally emerge due to hydrodynamic instabilities such as convective overturn and the standing accretion shock instability during the early moments following core collapse \citep{wongwathanarat13,wongwathanarat17, orlando21,orlando22}. These instabilities drive the expansion of plumes of Fe- and Ti-rich ejecta in an anisotropic manner, leading to a bulk offset in the ejecta distribution relative to the explosion center, as observed in X-rays by \citet{hwang04} and \citet{grefenstette14}. The resulting asymmetric explosion can also explain the motion of the central compact object, which moves southeast at $\approx 350$\kms \citep[P.A.$\approx170^\circ$; ][]{thorstensen01,hollandashford24,suzuki25}. However, the systemic LOS ejecta velocities predicted in such models are typically of order $\sim$100--200\kms, significantly smaller than the $\sim$800\kms\ offsets inferred for Cas~A. \citet{orlando22} further demonstrated that the interaction of the SNR with an asymmetric dense circumstellar shell, possibly ejected $10^4$--$10^5$~yr before core-collapse, can explain the observed inward motion of the reverse shock and enhanced synchrotron emission in the western hemisphere \citep[e.g., see][]{vink22}. In this picture, the reverse shock geometry and brightness asymmetries, as well as the asymmetry along the LOS, are reproduced, but the resulting geometric center of the shocked shell shifts eastward, opposite to the observed $\sim$0.2~pc displacement of Cas~A's ejecta shell toward the northwest \citep[see Figure~8 of ][]{orlando22}. Thus, although neutrino-driven explosion models coupled with a non-uniform circumstellar environment successfully reproduce many of Cas A’s observed asymmetries, the specific models explored to date do not fully account for the observed magnitude of the bulk motion. Exploring a wider range of explosion energies and CSM configurations, as suggested by \citet{orlando22}, will be necessary to assess whether the observed motion can be produced naturally within the neutrino-driven framework.

\subsection{Regional Variation of Expansion Rate \label{ssec:shelldecel}}

\begin{figure}
\fig{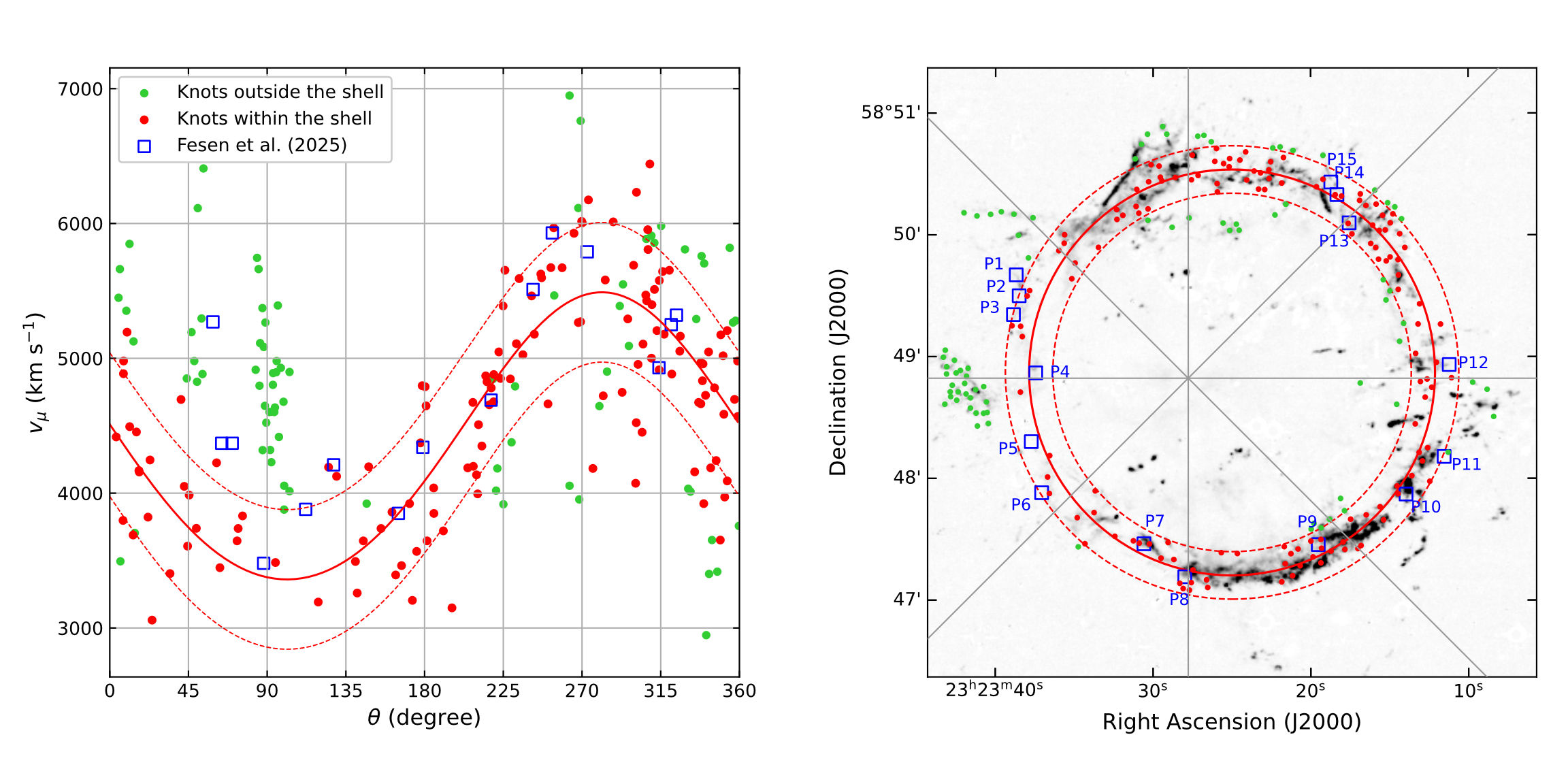}{0.9\textwidth}{}
\caption{\emph{Left:} Proper-motion velocities of main shell knots as a function of position angle. Red points indicate knots within the 2$\sigma$ range of the Gaussian ring model, while green points mark those outside the range. The red curve shows the uniformly-expanding, spherical-shell model from Figure~\ref{fig:muxmuy}, and the 2$\sigma$ bounds are shown by dashed lines. Blue boxes denote the shell ejecta motions reported by \citet{fesen25}. \emph{Right:} Sky positions corresponding to the components in the left panel. The background is the 2013 \img. Ejecta positions from \citet{fesen25} are labeled. Gray lines indicate the polar-angle grid used in the left panel. \label{fig:mpa}}
\end{figure}

Figure~\ref{fig:mpa} (left) presents the proper motion velocities of compact ejecta knots in the main shell as a function of position angle $\theta$ measured from the explosion center. Red symbols denote knots within the 2$\sigma$ range of the Gaussian ring model, while green points mark those outside the range (see the right frame). A pronounced east-west asymmetry in the expansion of the main shell knots is evident. Their distribution is well reproduced by a uniformly expanding shell model with a mean expansion velocity of $v_{\rm exp}=4425$\kms\ and a systemic motion of $(v_{\alpha,0}, v_{\delta,0})=(-1044, +210)$\kms\ (solid red curve). Individual ejecta knots within the shell, however, exhibit significant velocity scatter, bounded by the dashed lines that represent the expected 2$\sigma$ range from the Gaussian ring ($\pm 1035$\kms, dashed red curve), corresponding to the finite thickness of the shell. 

A notable feature in Figure~\ref{fig:mpa} (left) is the near absence of knots expanding with velocities less than the solid red line at P.A.$\simeq45^\circ$--135$^\circ$. This region corresponds to where the shell is disrupted, and the knots there are associated with filamentary structures extending beyond the nominal shell radius. These filaments likely represent fragments of the disrupted shell, and the trends in Figure~\ref{fig:mpa} suggest that this portion of the shell is expanding slightly faster than before disruption or that the outer portion has survived selectively.

The knots located outside the main shell (green symbols) show much larger scatter. In particular, those in the northeastern region (P.A.$\approx45^\circ$ and $90^\circ$) exhibit markedly higher velocities, reaching $\sim$6000\kms, about 1.5 times faster than the main shell. These fast knots coincide with the ``rupture'' or protrusion that extends beyond the nominal shell boundary (see the right frame in Figure~\ref{fig:mpa}). The rupture appears as a flared opening of the otherwise continuous ring, through which the interior ejecta escaped \citep[e.g.,][]{milisavljevic24}. In contrast, knots in the western quadrant exhibit a broader range of velocities, from $\sim$4000 to 7000\kms, indicating a significant spatial dispersion of knots in this region.

\begin{figure}
\fig{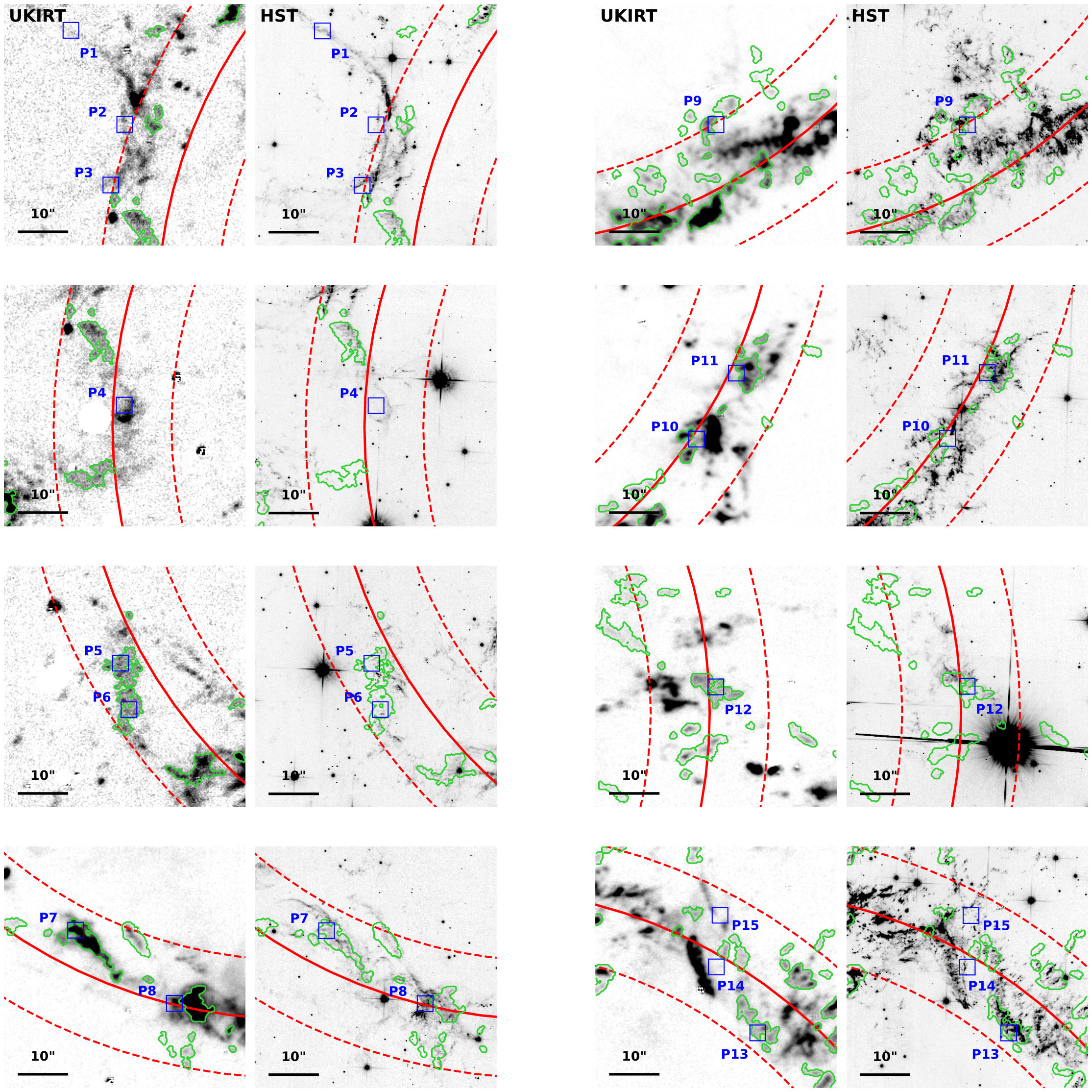}{0.9\textwidth}{}
\caption{Ejecta knots analyzed by \citet{fesen25}. Each panel pair shows a $0\farcm8\times0\farcm8$ cutout, with the 2013 \img\ on the left and the 2022 F775W \hst\ on the right. The \hst\ image has been shifted to the 2013 epoch, assuming free expansion. Blue squares mark the P1--P15 knot positions defined by \citet{fesen25}. Green contours indicate the regions from which proper motions are derived in this study (Section~\ref{sec:shell}). Red solid curves trace the Gaussian ring models shown in Figure~\ref{fig:mpa}. A 10\arcsec scale bar is plotted in the lower left of each panel. \label{fig:fesencomp}}
\end{figure}

In Figure~\ref{fig:mpa}, the small blue squares show the results of \citet{fesen25}, who measured proper motions of optical knots in the main shell using \hst\ data. Their measurements generally follow the spherical-shell expansion model, except at P.A.$\simeq 45^\circ$--$90^\circ$, where their data points (P1--P3 and P5--P6) lie substantially above the solid red line. These knots correspond to filamentary structures that likely represent fragments of the disrupted portion of the main shell, extending slightly beyond the nominal shell boundary. We directly compared our ejecta proper motion measurements with those listed in Table~5 of \citet{fesen25}. Among the fifteen features cataloged in their study, nine could be identified in our data set with one-to-one correspondence: P5, P6, P7, P8, P9, P10, P11, P12, and P13 (see Figure~\ref{fig:fesencomp}). Three features are not directly identified, but their adjacent features could be compared: P3, P14, and P15. The measurements for P5, P6, P10, P13, and P15 are consistent within 5\%, whereas P3, P7, P8, P9, P11 and P14 show $\sim$10\% slower motions and P12 shows $\sim$10\% faster motion. These discrepancies are likely attributable to local brightness variations and complex substructures in which multiple clumps are enclosed within a single contour, rather than to genuine differences in ejecta motion. The remaining three positions were too faint or had evolved significantly and thus could not be reliably recovered in our analysis.

\citet{fesen25} noted that knots in the southeastern part of the shell (P5--P8) expand significantly more slowly (3900--4300\kms) than those in the northeastern and northwestern shells (P1 and P13-P15; 4900--5300\kms). They interpreted this as evidence for a north-south or northwest-southeast asymmetry in the SN explosion, in which more ejecta mass was ejected at lower velocities toward the rear (redshifted) side of SNR, corresponding to its projected northern and western regions. This inferred northwest-southeast explosion asymmetry is consistent with the distribution of Fe-rich ejecta seen in X-rays and with the southeastward proper motion of the compact X-ray point source. It also replicates the long-recognized asymmetry in both optical and X-ray emission, where the northern limb of Cas~A has historically dominated the optical flux, while the southern regions have gradually brightened over the past five decades as the reverse shock advances into that area \citep[see ][ for an in-depth discussion]{fesen25}. According to our measurements, however, the primary velocity asymmetry is oriented nearly east-west rather than north-south. In the results of \citet{fesen25}, the western knots (P10--P12) also have the highest expansion velocities. Moreover, the shell's expansion properties are likely affected by the complex structure of the surrounding CSM \citep[e.g., ][]{orlando22}. For example, in the southern shell, the segment between P8 and P9 (P.A.$\approx 170^\circ$--$200^\circ$) is bright in [Fe II] emission but faint in optical ejecta emission, which may indicate the presence of a substantial amount of dense circumstellar material (\citealt{lee17}; see also \citealt{vink24}). Therefore, the connection between the observed asymmetries in the shell expansion and the intrinsic asymmetry of the SN explosion remains uncertain and warrants further investigation. 

\citet{fesen25} further estimated the velocities of the reverse shock by identifying newly brightened ejecta knots that mark the advance of the reverse shock through the main shell. The inferred reverse-shock velocities in the sky frame typically range 1000--2000\kms, about 1000\kms\ lower than those obtained from X-ray measurements. They also reported that the reverse shock is nearly stationary along the western limb, though not inward moving at velocities as large as $\sim$2000\kms\ as previously suggested from X-ray studies \citep{helder08,sato18,vink22}. We examined the 2013 and 2020 \img\ to search for similar signatures of reverse-shock motion. However, due to the low surface brightness, large seeing size, and limited temporal baseline, it is difficult to distinguish between deceleration and reverse-shock motion in our data.

\section{Conclusion and Summary \label{sec:conclusion}}

Cas~A is a young core-collapse SNR whose extent and morphology have continued to evolve since its discovery in the 1950s. Early optical observations showed an incomplete ejecta structure, with the southern portion of the main shell being faint or entirely absent. By the 1980s, however, this region had brightened considerably, and Cas~A now appears as a nearly complete, circular ring of emission encompassing the entire SNR. In this study, we conducted a detailed comparative analysis of Cas~A using two deep, narrow-band \fe+\si\ (1.644~\micron) images obtained with UKIRT/WFCAM in 2013 and 2020. The 2020 observations were performed using the same instrument configuration and reduction procedures as those employed for the 2013 dataset \citep{koo18}, allowing a direct high-precision comparison of the two epochs. A difference image constructed from the two revealed the characteristic white-black pairing pattern that traces the outward motion of the shocked ejecta in the plane of the sky. In these images, the QSFs and the rapidly expanding ejecta --- comprising the main shell and the FMKs --- are clearly distinguished, enabling us to track their brightness variations, morphological evolution, and proper motions over the seven-year baseline. We produced a catalog of compact sources, including both QSFs and FMKs, and investigated their physical and kinematic properties. Our main findings are summarized below.

\begin{enumerate}

\item 
A total of 263 compact features were identified, comprising 132 QSFs and 131 FMKs (Table~\ref{tab:catalog}). Among them, 39 knots were newly cataloged in 2020. In contrast, 85 knots listed in the 2013 catalog of \citet{koo18} are no longer detected, most of which were FMKs. All knots whose classifications were uncertain and listed as QSF or FMK candidates in the previous study are now fully identified.

\item
The total \fe\ flux from QSFs in 2020 was found to be nearly identical to the 2013 value, but some individual QSFs show significant flux variations, particularly the faint ones. The QSFs along the southern arc structure generally brightened, while the QSFs in the interior and western regions generally darkened. For comparison, the total flux of FMKs slightly increased from 2013. Individual FMKs exhibit greater variations than in QSFs, in some cases by an order of magnitude. 

\item
We measured the proper motions of the outer FMKs using cross-correlation and derived expansion indices $m$, defined as $m \equiv d\ln r/d\ln t$. The knots show an overall tendency toward ballistic expansion with $m\approx 1.0$. However, a systematic deceleration is evident in the Fe~K~plume area ($\theta =78^\circ$--$150^\circ$), where the mean $m$ value is $0.964\pm0.010$. Knots in the northern area exhibit modest deceleration ($m=0.974$) with somewhat larger uncertainties. These results suggest that dense FMKs have undergone only moderate slowing as they crossed the reverse shock, in contrast to the surrounding diffuse X-ray ejecta, which shows substantially stronger deceleration.

\item
We also measured the proper motions of ejecta knots in the main shell to examine its global expansion. The shell exhibits a pronounced east-west asymmetry in its expansion, with its geometric center offset by $\sim 22\arcsec$ ($\sim 0.4$ pc) toward the northwest relative to the explosion center. These spatial and kinematic characteristics are well described by a uniformly expanding shell with a systemic motion. The shell expands at $0\farcs275$ yr$^{-1}$ (4400\kms) and shows a systemic motion of $0\farcs066$ yr$^{-1}$ (1065\kms) toward the northwest (Table~\ref{tab:shellparameter}). However, individual knots display substantial velocity scatter, showing regional variations across the shell. These findings are consistent with previous studies. In three dimensions, the shell moves toward the northwest in the sky plane and away from the observer along the LOS, with a systemic velocity of $\sim1300$\kms.

\end{enumerate}

\begin{acknowledgements}
 We thank Salvatore Orlando and Jacco Vink for their helpful and insightful comments on the manuscript. We also thank the staff of the United Kingdom Infrared Telescope (UKIRT) for their support. The data used in this study were obtained under the UKIRT Service program [Project ID: U/18B/EAP003], and we are grateful to the operators and technical staff for their assistance with the observations. When some of the data reported here were obtained, the operations were enabled through the cooperation of the East Asian Observatory. UKIRT is owned by the University of Hawaii (UH) and operated by the UH Institute for Astronomy. The JWST/NIRCam images shown in Figure~2 were produced using data obtained as part of the Cycle~1 program GO-01947 (PI: D. Milisavljevic), acquired in 2022 August and November. All the HST and JWST data used in this paper can be found in MAST: \dataset[10.17909/at5s-qj50]{http://dx.doi.org/10.17909/at5s-qj50}. This research was supported by the Basic Science Research Program through the NRF of Korea, funded by the Ministry of Science, ICT and Future Planning (RS-2023-00277370). 
\end{acknowledgements}

\vspace{5mm}

\facilities{Chandra X-ray Observatory ({\textrm{Chandra}}), Hubble Space Telescope (\hst), James Webb Space Telescope (\jwst), Karl G. Jansky Very Large Array (VLA), United Kingdom Infrared Telescope (UKIRT)}
\software{{Astropy \citep{astropy13,astropy18,astropy22}}, DS9 fitsviewer \citep{joye03}, IDL, image-registration, Matplotlib\citep{hunter07}, NumPy\citep{harris20}, photutils\citep{bradley25}, SciPy\citep{virtanen20}, SWARP \citep{bertin02} }

\end{CJK*}
\end{document}